\documentstyle[preprint,aps,eqsecnum]{revtex}
\tighten

\begin{document}

\preprint{NIKHEF-94-P1, hep-ph/9403227}

\draft

\title{Intrinsic transverse momentum\\ and the polarized Drell-Yan process}

\author{R.D. Tangerman and P.J. Mulders\thanks{Also at Physics Department,
Free University, NL-1081~HV Amsterdam, The Netherlands.}}

\address{National Institute for Nuclear Physics and High Energy Physics
(NIKHEF-K),\\ P.O. Box 41882, NL-1009~DB Amsterdam, The Netherlands }

\date{March 1994, revised August 1994}

\maketitle

\begin{abstract}
In this paper we study the cross section at leading order in $1/Q$ for
polarized Drell-Yan scattering at measured lepton-pair transverse
momentum $Q_T$.
We find that for a hadron with spin $1/2$ the quark content at leading order
is described by six distribution functions for each flavor, which depend
on both the lightcone momentum fraction $x$, and the quark transverse
momentum $\bbox{k}_T^2$.
These functions are illustrated for a free-quark ensemble.
The cross sections for both longitudinal and transverse polarizations are
expressed in terms of convolution integrals over the distribution functions.
\end{abstract}


\section{Introduction}

The measurements of unpolarized structure functions in deep inelastic
scattering (DIS) of leptons off nucleons and nuclei and those of polarized
structure functions in scattering of longitudinally polarized electrons
off longitudinally polarized nucleons ~\cite{ESMC} have yielded
the lightcone momentum distributions
$f_1(x)$ for quarks in various targets and the helicity distributions
$g_1(x)$ in protons and neutrons\footnote{Another often used notation
is $q(x)$ for the lightcone momentum distribution and $\Delta q(x)$ for
the helicity distribution ($q=u,d,s,\ldots$).}.
These measurements, and particularly their interpretation, have shown the
importance of understanding the relation of these  distributions to
the structure of the target. The distributions $f_1(x)$ and $g_1(x)$
characterize the response of the hadron in inclusive DIS at leading
order in the  transferred momentum $Q$.
In inclusive deep inelastic lepton-hadron ($\ell H$) scattering the
quark transverse momentum is not observable, since it is integrated over.
In the Drell-Yan (DY) process at measured lepton-pair transverse momentum
$Q_T$, however, quark transverse momentum {\em does\/} enter in observables,
notably in the angular distribution of the lepton pairs.
The main point of this paper is the discussion of quark
transverse momentum in polarized Drell-Yan scattering. We will restrict
ourselves to leading order and discard contributions
which are suppressed by orders of $1/Q$. We will also not discuss QCD
radiative corrections, giving rise to logarithmic corrections.

For inclusive deep inelastic $\ell H$ scattering, assuming only one flavor,
the hadron tensor is given as the imaginary part of the forward virtual Compton
amplitude, for large virtual photon momentum $q$ ($Q^2\equiv -q^2$ large) given
by the sum of the quark and antiquark handbag diagrams of
Fig.~\ref{fig:handbag}. The basic object, encoding
the soft physics of the quarks inside the hadron, is the correlation function%
{}~\cite{sope,jaff83}
\begin{equation}
\Phi_{ij}(P S;k) =
\int \frac{ d^4 x }{(2\pi)^4}\; e^{i k\cdot x}
\langle P S | \overline{\psi}_j(0) \psi_i(x)|P S\rangle_c , \label{eq01}
\end{equation}
where $k$ is the momentum of the quark. The vectors $P$ and $S$
are the momentum and spin vector of the target hadron. Evaluating the hard
part,
the scattering of the virtual photon off the quarks, it turns out that the
structure functions in the cross section become proportional to
$f_1(x_{\text{bj}})$ and $g_1(x_{\text{bj}})$, where
$x_{\text{bj}} = Q^2/2P\cdot q$. The function $f_1$ is given by
\begin{equation}\label{feenx}
f_1(x) = \frac{1}{2}\int d k^- d^2  \bbox{k}_T\;\text{Tr}
\left[\gamma^+\Phi(P S;k)\right] ,
\end{equation}
where $x=k^+/P^+$. It can be interpreted as the longitudinal (lightcone)
momentum distribution of quarks. The function $g_1$ appears as
\begin{equation}\label{geenx}
\lambda \,g_1(x) = \frac{1}{2}\int d k^- d^2
\bbox{k}_T\;\text{Tr}\left[\gamma^+\gamma_5\Phi(P S;k)\right],
\end{equation}
and can be interpreted as the quark helicity distribution in a longitudinally
polarized nucleon (helicity $\lambda= 1$). The functions $f_1$ and $g_1$ are
specific projections of $\Phi$. Which projections of $\Phi$ contribute in hard
scattering processes in leading order can be investigated by looking at the
operator structure, including the
Dirac and Lorentz structure, of the correlation function. Such an analysis
requires some physical constraints on the range of quark momenta.
The analysis of $\Phi$ (integrated over $k^-$ and $\bbox{k}_T$) shows that
there is one more leading function, the transverse polarization or
transversity\footnote{The authors of~\cite{jaji} use the name `transversity
distribution' in order to make clear that a quark
of definite transversity is {\em not\/} in an eigenstate of the transverse
spin operator but of the Pauli-Lubanski operator projected along a transverse
direction. The authors of\cite{cort92} object to this nomenclature,
because of the pre-existence of the term,
and prefer to call it transverse polarization distribution.}
distribution $h_1$. It is related to a bilocal quark-quark matrix element
through~\cite{rals79}
\begin{equation}\label{heenx}
\bbox{S}_{T}^i\,h_1(x) = \frac{1}{2}\int d k^- d^2 \bbox{k}_T\;
\text{Tr}\left[i\sigma^{i+}\gamma_5\Phi(P S;k)\right]\qquad\qquad (i=1,2),
\end{equation}
which shows that $h_1$ can be interpreted as the quark transversity
distribution in a transversely polarized nucleon.
This is a chiral-odd distribution,
which is not observable in inclusive $\ell H$ scattering. It needs to be
combined with some other chiral-odd structure, e.g., the fragmentation part in
semi-inclusive leptoproduction of hadrons or the antiquark distribution part
of DY scattering\cite{jaji,cort92,artr90,xiji92}.

In this paper we discuss one possible way to extract more information from the
correlation function $\Phi$. We are after the dependence on the transverse
momentum $\bbox{k}_T$. One way to study this dependence is the observation of
a hadron in the outgoing quark jet, e.g., in semi-inclusive $\ell H$
scattering~\cite{leve94}. This process, however, also requires consideration
of the fragmentation functions. In this paper we study the
process that is sensitive to intrinsic transverse momentum and involves only
quark distribution functions, namely, massive dilepton production or the
Drell-Yan (DY) process~\cite{drya}.

About fifteen years ago Ralston and Soper (RS) published a pioneering
paper~\cite{rals79} on the polarized Drell-Yan process.
Because we take it as our starting point, we
briefly sketch its content. RS write down a covariant expansion for
$\int dk^- \Phi$,
which is the quantity that is relevant in the hadron tensor for the
DY process. To determine this expansion they use symmetry arguments and an
infinite-momentum-frame analysis. They find five independent distribution
functions, divided in one momentum probability distribution ${\cal
P}(x,\bbox{k}_T^2)$, two functions describing the quark helicity, and
two describing its transverse polarization. With these they calculate
the polarized Drell-Yan cross section with the virtual photon transverse
momentum $Q_T\equiv\sqrt{\bbox{q}_{T}^2}$ put to zero. In that case they
are sensitive to four of the five distribution functions.
When they integrate over the transverse momentum
they are only sensitive to three distribution functions.

We extend on these results in two ways. First, we show that RS left out one
transverse momentum distribution, needed to describe the quark transverse
polarization. This additional function is obtained using general symmetry
arguments. It also shows up in a model that we are going to
employ later and that describes a gas of free partons.
Our second extension is the calculation of the polarized DY cross section
without constraints on $Q_T$ [other than it being of ${\cal O}(\Lambda)$],
thereby becoming sensitive to all six distribution functions.

We end this introduction with a remark on possible QCD corrections affecting
transverse momenta and factorization.
A difficulty of the extra scale $Q_T$ is the Sudakov effect. Soft gluon
radiation gives rise to radiative transverse momentum. However, the large
logarithms  connected with this effect can be summed and exponentiated to
Sudakov form  factors~\cite{CSS85}. From these it becomes clear that if
$Q_T$ is sufficiently low, i.e., of hadronic  scale $\Lambda$, as compared
to $Q$, the transverse momentum governing the process is predominantly
intrinsic.
Factorization means that the process can be written as a convolution of
renormalized distribution functions and a perturbatively calculable
short-distance part. For polarized DY at measured $Q_T\lesssim\Lambda$
factorization has not been proven  yet~\cite{coll93a,CSS89}. We will not
further address this problem here, but use the diagrammatic expansion
proposed by Ellis, Furma\'nski, and Petronzio (EFP)~\cite{EFP}
to study the DY process. In this diagrammatic expansion Green functions appear,
incorporating the long-range QCD physics.
These correlation functions are connected by ordinary Feynman graphs with
quarks and gluons, the hard scattering piece.

The outline of this paper is as follows. In Sec.~\ref{DYp} we give the
one-photon exchange picture for massive dilepton production. We specify the
notation in a frame where the two hadrons are collinear, and the axes are
given with respect to which the lepton angles are defined.
In Sec.~\ref{For} we analyze the quark correlation
function, and find six leading distribution functions.
In Sec.~\ref{Fqm} we discuss the free-quark ensemble as
an example. In Sec.~\ref{Res} we calculate the leading-order hadron tensor
and present cross sections for various combinations of polarizations.
We end with a discussion of these results.

\section{The Drell-Yan process} 			\label{DYp}

In this section we want to discuss the cross section, kinematic aspects,
and structure functions, for polarized Drell-Yan scattering.
For a complete overview we refer for the unpolarized process to
Lam and Tung~\cite{lam78}, and for polarized DY scattering to
Donohue and Gottlieb~\cite{dono81}, who make use of
the Jackob-Wick helicity formalism.

\subsection{The DY cross section} 		        \label{gcs}

We consider  the process $A+B\rightarrow \ell
+\bar{\ell}+X$, where two spin-$\frac{1}{2}$ hadrons with
momenta $P_A^\mu$ and $P_B^\mu$ interact and two outgoing leptons
are measured with momenta $k_1^\mu$ and $k_2^{\mu}$.
The leptons are assumed to originate from a high-mass photon with momentum
$q = k_1 + k_2$, with $Q^2\equiv q^2 > 0$.
We consider the case of pure incoming spin states, characterized by the
spin vectors $S_A^\mu$ and $S_B^\mu$, i.e., $S_A^2=S_B^2=-1$.
In the deep inelastic limit $Q^2$ and $s=(P_A + P_B)^2$ become large
compared to the characteristic hadronic scale of order $\Lambda^2
\sim 0.1 \text{\ GeV}^2$, while their ratio $\tau=Q^2/s$ is fixed.
The phase space element for the lepton pair can be written as $d^4q\,d\Omega$,
where the angles are those of the lepton axis in the dilepton rest
frame with respect to a suitably chosen Cartesian set of axes.
The cross section can be written as
\begin{equation}\label{cross}
\frac{d\sigma}{d^4 q d\Omega}=\frac{\alpha^2}{2s\,Q^4} L_{\mu\nu}W^{\mu\nu},
\end{equation}
where the lepton tensor is given by (neglecting the lepton masses)
\begin{equation}\label{leptten}
L^{\mu\nu} =
2\,k_1^{\mu}k_2^{\nu}+ 2\,k_2^{\mu}k_1^{\nu}- Q^2\,g^{\mu\nu}  ,
\end{equation}
and the hadron tensor can be written as
\begin{equation}
W^{\mu\nu}( q; P_A S_A; P_B S_B )=
\int \frac{d^4 x}{(2\pi)^4}\; e^{i q\cdot x} \langle  P_A S_A; P_B S_B |\,
[J^\mu (0), J^\nu (x)]\, | P_A S_A; P_B S_B \rangle .  \label{hadrten}
\end{equation}
Since the lepton tensor~(\ref{leptten}) is symmetric in its indices, we will
from now on only consider the symmetric part of $W^{\mu\nu}$.

\subsection{Kinematics}

We define the transverse momentum
of the produced lepton pair in a frame where the hadrons
are collinear, with the third axis chosen along the direction of hadron $A$.
One has $\bbox{q}_{T}^2\equiv Q_T^2\lesssim \Lambda^2$.
It is convenient to work in a lightcone component
representation, $p=[ p^-,p^+,\bbox{p}_{T} ]$ with
$p^\pm \equiv (p^0 \pm p^3)/\sqrt{2}$. The momenta of
the hadrons and the virtual photon in a collinear frame take the form
\begin{eqnarray}
&& P_A = \left[ \frac{M_A^2}{2P_A^+},  P_A^+ , \bbox{0}_{T} \right]
\approx \left[ \frac{x_A M_A^2}{\sqrt{2}\,\kappa Q},\frac{\kappa  Q}
{\sqrt{2}\,x_A},\bbox{0}_{T}\right], \label{pamom}\\
&& P_B = \left[ P_B^-, \frac{M_B^2}{2P_B^-} ,\bbox{0}_{T} \right]
\approx \left[ \frac{Q}{\sqrt{2}\,\kappa x_B},\frac{\kappa x_B M_B^2}
{\sqrt{2}\, Q},\bbox{0}_{T}\right],  \label{pbmom}\\
&& q = \left[ x_B P_B^-,  x_A P_A^+ , \bbox{q}_{T} \right]
\approx \left[ \frac{Q}{\sqrt{2}\,\kappa},\frac{\kappa Q}{\sqrt{2}},
\bbox{q}_{T}\right],                   \label{qmomentum}
\end{eqnarray}
neglecting corrections of order $1/Q^2$, indicated here and further on
by an approximate equal. The parameter $\kappa$ fixes the collinear frame.
One has $\kappa=x_A M_A/Q$ for the frame in which hadron $A$ is at rest,
$\kappa=\sqrt{x_A/x_B}$ for the hadron center-of-mass frame, and
$\kappa = Q/x_BM_B$ for the frame in which hadron $B$ is at rest.
The following Lorentz-invariant relations hold:
\begin{eqnarray}
&& x_A = \frac{q^+}{P_A^+} \approx \frac{Q^2}{2P_A\cdot q}
\approx \frac{P_B\cdot q}{P_B \cdot P_A}, \\
&& x_B = \frac{q^-}{P_B^-} \approx \frac{Q^2}{2P_B\cdot q}
\approx \frac{P_A\cdot q}{P_A \cdot P_B},\\
&& s \approx 2 P_A^+ P_B^- \approx \frac{Q^2}{x_A x_B} .
\end{eqnarray}
The above relations also show that all dot products for any pair from
the vectors $q$, $P_A$, and $P_B$, are of order $Q^2$. As compared to this,
the hadron momenta are almost lightlike. We can define the exactly
lightlike vectors that in a given collinear frame have the form
\begin{eqnarray}
& & n_+ \equiv [ 0, \kappa ,\bbox{0}_{T}], \nonumber\\
& & n_- \equiv [ \kappa^{-1} , 0, \bbox{0}_{T}],  \label{nanb}
\end{eqnarray}
satisfying $n_+ \cdot n_-= 1$.
Given an arbitrary four-vector $a$, and the projector
\begin{equation}\label{getrans}
g_T^{\mu\nu}\equiv  g^{\mu\nu} -n_+^{\mu} n_-^{\nu}- n_+^{\nu} n_-^{\mu} ,
\end{equation}
we define the spacelike {\em transverse\/} four-vector
$a_T^\mu\equiv g_T^{\mu\nu}a_\nu$, or, in coordinates in a collinear
frame, $a_T=\left[0,0,\bbox{a}_{T}\right]$.
Note that for any transverse vector one has
\begin{equation}
a_T\cdot P_A=a_T\cdot P_B=0.
\end{equation}

For the analysis of the hadronic tensor which satisfies
$q^\mu\,W_{\mu \nu}=q^\nu\,W_{\mu \nu}= 0$, it is important to construct
vectors that are orthogonal to $q$. We will use the projector
\begin{equation}
\tilde g^{\mu \nu} \equiv g^{\mu\nu}- \frac{q^\mu q^\nu}{q^2}
\end{equation}
for this, and define
\begin{equation}
\tilde a^\mu\equiv\tilde g^{\mu \nu} a_\nu=a^\mu -\frac{a\cdot q}{q^2}\,q^\mu.
\end{equation}
As $q$ is timelike in DY scattering, it is useful to define a set of
Cartesian axes. The $Z$-direction, known as the Collins-Soper
axis~\cite{coll77}, is chosen as in~\cite{rals79}, but our
$X$- and $Y$-direction are opposite.
To be precise, we use ($\epsilon^{0123}=1$)
\begin{eqnarray}
Z^\mu & \equiv &  \frac{P_B\cdot q}{P_B\cdot P_A}\,\tilde P^\mu_A
              -\frac{P_A\cdot q}{P_A\cdot P_B}\,\tilde P^\mu_B
           \ =\   \frac{P_B\cdot q}{P_B\cdot P_A}\, P^\mu_A
              -\frac{P_A\cdot q}{P_A\cdot P_B}\, P^\mu_B, \nonumber\\
X^\mu & \equiv & -\frac{P_B\cdot Z}{P_B\cdot P_A}\, \tilde P^\mu_A
             +  \frac{P_A\cdot Z}{P_A\cdot P_B}\, \tilde P^\mu_B,
\label{assen} \\
Y^\mu & \equiv & \frac{1}{P_A\cdot P_B}\,\epsilon^{\mu \nu \rho \sigma}
P_{A\nu} P_{B\rho} q_\sigma .\nonumber
\end{eqnarray}
These vectors are orthogonal and satisfy $Z^2\approx -Q^2$
and $X^2\approx Y^2 \approx q_T^2=-Q_T^2$.
They form a natural set of spacelike axes (with in the dilepton rest frame
only spatial components). We will denote $\hat q^\mu=q^\mu/Q$,
$\hat z^\mu=Z^\mu/\sqrt{-Z^2}$, etcetera.
Explicitly, one has in a collinear frame
\begin{eqnarray}
\hat{q}&\approx&\left[ \frac{1}{\sqrt{2}\,\kappa} ,\frac{\kappa}{\sqrt{2}},
\frac{\bbox{q}_{T}}{Q}\right],\nonumber\\
\hat{z}&\approx&\left[ -\frac{1}{\sqrt{2}\,\kappa},\frac{\kappa}{\sqrt{2}},
\bbox{0}_{T}\right],\nonumber\\
\hat{x}&\approx&\left[ \frac{1}{\sqrt{2}\,\kappa}\,\frac{Q_T}{Q} ,
\frac{\kappa}{\sqrt{2}}\,\frac{Q_T}{Q} ,  \label{qzxy}
\frac{\bbox{q}_{T}}{Q_T}\right],\label{ascoor}\\
\hat{y}&\approx&\left[ 0,0,\frac{\bbox{y}_{T}}{Q_T}\right],\nonumber
\end{eqnarray}
where $\bbox{y}_{T}^i=\epsilon^{ij}\bbox{q}_{T j}$.
Note that since $Z^\mu$ is a linear combination of the hadron momenta,
it has in collinear frames no transverse components.
The transverse vectors $a_T$, thus, are orthogonal to $Z$. They are,
in general, not orthogonal to $q$. One has, for example,
$q_T^\mu\approx X^\mu - (Q_T^2/Q^2)\,q^\mu$.
Note that the second term is only order $1/Q$ suppressed.
For an arbitrary four-vector $a$ we define the {\em perpendicular\/}
four-vector $a_\perp$ as the projection of the transverse vector $a_T$,
using the projector
\begin{equation}\label{geperp}
g_\perp^{\mu\nu}\equiv g^{\mu\nu}-\hat{q}^{\mu}\hat{q}^{\nu}
+\hat{z}^{\mu}\hat{z}^{\nu} ,
\end{equation}
yielding
\begin{equation}
a_\perp^\mu \equiv g_\perp^{\mu \nu}a_{T \nu}
=a_T^\mu - \frac{q\cdot a_T}{q^2}\,q^\mu.
\end{equation}
Thus, any perpendicular vector satisfies
\begin{equation}\label{perpcond}
a_\perp\cdot \hat{q}=a_\perp\cdot \hat{z}=0.
\end{equation}
Note that $X$ is the perpendicular projection of $q$.
The vectors $n_+$ and $n_-$, defined in Eq.~(\ref{nanb}),
can be expressed in terms of the set~(\ref{qzxy}),
\begin{eqnarray}
n_+ \approx \frac{1}{\sqrt{2}}
\left(\hat{q} + \hat{z} -\frac{Q_T}{Q}\,\hat{x} \right), \nonumber\\
n_- \approx \frac{1}{\sqrt{2}}
\left(\hat{q} - \hat{z} -\frac{Q_T}{Q}\,\hat{x} \right),
\end{eqnarray}
Inserting these into the definitions of the projectors $g_\perp$ and $g_T$,
one derives the relation
\begin{equation}\label{Pmetric}
g_\perp^{\mu\nu}\approx g_T^{\mu\nu}-
\frac{(\hat{q}^{\mu} q_T^{\nu}+ \hat{q}^{\nu} q_T^{\mu}) }{Q},
\end{equation}
from which one obtains for a general vector $a$:
\begin{equation}\label{Pvector}
a_\perp^\mu \approx a_T^\mu - \frac{a_T\cdot q_T}{Q}\,\hat{q}^\mu,
\end{equation}
provided that $a_T\cdot q_T={\cal O}(1)$.
{}From this expression it is evident that for two arbitrary vectors
$a$ and $b$, satisfying this condition,
\begin{equation}\label{Pinprod}
a_\perp\cdot b_\perp\approx a_T\cdot b_T .
\end{equation}
Restricting oneself to leading order, the vectors $a_T$ and $a_\perp$ can be
freely interchanged. In a higher order study, however, the difference will
become important~\cite{tang94}.

For the spin vectors the above definitions can be illustrated.
In a collinear frame the spin vectors, satisfying $P_A\cdot S_A
 = P_B\cdot S_B = 0$, can be written as
\begin{eqnarray}
S_A&=&\left[ -\lambda_A\frac{M_A}{2P_A^+},\lambda_A \frac{ P_A^+ }{M_A},
\bbox{S}_{AT}\right], \label{samom}\\
S_B&=&\left[\lambda_B \frac{P_B^-}{M_B}  ,-\lambda_B \frac{ M_B}{2P_B^-} ,
\bbox{S}_{BT}\right]. \label{sbmom}
\end{eqnarray}
where $\lambda_A$, and $\lambda_B$, are the hadron helicities.
The two-component vectors $\bbox{S}_{AT}$, and $\bbox{S}_{BT}$, give the
transverse polarization. Since we consider pure spin states, they obey
$\lambda^2 +\bbox{S}_{T}^2 = 1$. For the spin vectors we
have in a collinear frame $S_{T}=\left[ 0,0,\bbox{S}_{T}\right]$.
The perpendicular spin vector is given by
\begin{equation}
S_{\perp}^\mu \approx S_{T}^\mu - \frac{S_T \cdot q_T}{Q}\,\hat{q}^\mu
= \left[ {\cal O}\left(\frac{1}{\kappa Q}\right),{\cal O}
\left(\frac{\kappa}{Q}\right), \bbox{S}_{T}+{\cal O}\left(\frac{1}{Q^2}\right)
\right],
\end{equation}
where the longitudinal components follow from the transverse components by
demanding Eq.~(\ref{perpcond}), and using Eq.~(\ref{ascoor}).
If the spin vector would have been projected directly onto the $XY$-plane with
$g_\perp^{\mu \nu}$, one would have got
\begin{equation}
g_\perp^{\mu \nu} S_{\nu} = \left[ {\cal O}\left(\frac{1}{\kappa Q}\right),
{\cal O}\left(\frac{\kappa}{Q}\right), \bbox{S}_{T} - \frac{\lambda}{2\,xM}
\,\bbox{q}_{T}+{\cal O}\left(\frac{1}{Q^2}\right)\right].
\end{equation}
This differs from $S_{T}$ in the transverse sector by ${\cal O}(1)$,
unless $Q_T=0$.

\subsection{Structure functions} 			\label{sf}

With the definition~(\ref{assen}) of a Cartesian set of vectors orthogonal to
$q$, we can expand the lepton momenta in the following way:
\begin{eqnarray}
& & k_1^\mu =\case{1}{2} q^\mu +\case{1}{2} Q (\sin\theta\cos\phi\;\hat{x}^\mu
+ \sin\theta\sin\phi \; \hat{y}^\mu +\cos\theta \; \hat{z}^\mu ),\nonumber\\
& & k_2^\mu =  \case{1}{2} q^\mu - \case{1}{2} Q (\sin\theta\cos\phi \;
\hat{x}^\mu + \sin\theta\sin\phi \; \hat{y}^\mu +\cos\theta \; \hat{z}^\mu ).
\end{eqnarray}
Inserting these into Eq.~(\ref{leptten}), and using some trivial
goniometric relations and the completeness relation
$g^{\mu\nu}=\hat{q}^\mu\hat{q}^\nu-\hat{z}^\mu\hat{z}^\nu-
\hat{x}^\mu\hat{x}^\nu-\hat{y}^\mu\hat{y}^\nu$, we obtain
\begin{eqnarray}
L^{\mu\nu}= -\frac{Q^2}{2}&&\Bigl[ (1+\cos^2\theta)
\; g_\perp^{\mu\nu} -2\,\sin^2\theta\; \hat{z}^\mu \hat{z}^\nu
+ 2\sin^2\theta\cos2\phi
\; (\hat{x}^\mu \hat{x}^\nu + \case{1}{2}g_\perp^{\mu \nu}) \nonumber\\
& & + \sin^2\theta\sin 2\phi\; \hat{x}^{\{\mu} \hat{y}^{\nu\} } +
 \sin 2\theta\cos \phi\; \hat{z}^{\{\mu} \hat{x}^{\nu\} }+
\sin 2\theta\sin \phi\; \hat{z}^{\{\mu} \hat{y}^{\nu\} }\Bigr] ,
\label{hoeklep}
\end{eqnarray}
where the symmetrization of indices, $\hat{z}^{\{\mu} \hat{x}^{\nu\} }
\equiv \hat{z}^{\mu} \hat{x}^{\nu}+\hat{z}^{\nu} \hat{x}^{\mu}$,
is used. The six tensor combinations in Eq.~(\ref{hoeklep}) are not only
orthogonal to $q$, ensuring $q_\mu L^{\mu\nu}=0$, but also to each other.

It is convenient also to write the hadron tensor as a sum of products of
tensors and scalar functions, called {\em structure functions}.
{}From the properties of the electromagnetic current, one deduces for the
hadronic tensor [Eq.~(\ref{hadrten})] the following conditions:
\begin{equation}\label{hadcond}
\begin{array}{lcl}
q_\mu W^{\mu\nu}=0&\qquad&[\text{Current conservation}]\\
\left[W^{\nu\mu}\right]^*=W^{\mu\nu}&&[\text{Hermiticity}] \\
W_{\mu\nu}( \bar{q}; \bar{P}_A\, -\!\!\bar{S}_A; \bar{P}_B
\, -\!\!\bar{S}_B )=W^{\mu\nu}( q; P_A S_A; P_B S_B )&&[\text{Parity}]\\
W_{\mu\nu}( \bar{q}; \bar{P}_A\bar{S}_A;
\bar{P}_B \bar{S}_B )=\left[W^{\mu\nu}( q; P_A S_A; P_B S_B )\right]^*&&
[\text{Time reversal}]
\end{array}
\end{equation}
where $\bar{a}^\mu\equiv a_\mu$. The hermiticity condition, for instance,
requires that the symmetric part of $W^{\mu \nu}$ is real.
In unpolarized scattering the constraints imply the expansion
\begin{equation}\label{unpol}
W^{\mu\nu}= -(W_{0,0}-\case{1}{3}W_{2,0})\; g_\perp^{\mu\nu}
+ (W_{0,0}+\case{2}{3}W_{2,0})\;\hat{z}^\mu \hat{z}^\nu
- W_{2,1}\; \hat{z}^{\{ \mu} \hat{x}^{\nu \}}
-W_{2,2}\;(\hat{x}^\mu \hat{x}^\nu+\case{1}{2}g_\perp^{\mu \nu}),
\end{equation}
where the four structure functions depend on the (four) independent scalars,
or equivalently on $Q$, $x_A$, $x_B$, and $Q_T$. Since we choose to work
with the normalized vectors, the structure functions $W_{2,1}$ and $W_{2,2}$
contain kinematical zeros for $X^2\approx -Q_T^2=0$ of first and second order,
respectively. In that they differ from the ones in RS~\cite[Eq. (2.5)]{rals79}.
To be precise: our $W_{2,1}$ is $\sqrt{X^2 Z^2}$ times theirs,
and our $W_{2,2}$ is $-X^2$ times theirs. The linear combinations multiplying
$-g_\perp^{\mu\nu}$ and $\hat{z}^\mu\hat{z}^\nu$ are often referred to as
$W_T$ and $W_L$, respectively.
Inserting Eqs.~(\ref{hoeklep}) and~(\ref{unpol}) into Eq.~(\ref{cross}),
one has for unpolarized Drell-Yan scattering
\begin{equation}
\frac{d\sigma}{d^4 q d\Omega}=\frac{\alpha^2}{2s\,Q^2}\left[
2W_{0,0}+W_{2,0}\,(\case{1}{3}-\cos^2\theta)+W_{2,1}\sin 2\theta\cos\phi
+W_{2,2}\,\case{1}{2}\sin^2\theta\cos2 \phi\right].
\end{equation}
Due to the extra pseudo-vectors $S_A$ and $S_B$, in polarized
Drell-Yan there are several more structure functions.
We will not give them in general. Later we will simply consider
the ones that arise at leading order in $1/Q$ in the cross section.

\section{Formalism}					\label{For}

\subsection{The correlation function}

In order to calculate the hadron tensor, the diagrammatic expansion of EFP
is used~\cite{EFP}.
Each diagram is composed of soft nonlocal
matrix elements, convoluted with a hard cut-amplitude.
For example, in DY scattering the simplest diagrams (Born
diagrams) are given in Fig.~\ref{fig:DY}. The nonlocal matrix
element describing the nonperturbative long-range physics
for a quark of flavor $a$ inside hadron $A$, is given by the quark-quark
correlation function
\begin{equation}\label{corelgen}
(\Phi_{a/A})_{ij}( P_A S_A;k)=\int \frac{d^4 x}{(2\pi)^4}\; e^{i k\cdot x}
\langle P_A S_A |\overline{\psi}_j^{(a)}(0)\psi_i^{(a)}(x)|P_A S_A\rangle_c ,
\end{equation}
diagrammatically represented in
Fig.~\ref{fig:blob}. We will suppress the quark label $a$, the hadron
label $A$, and the connectedness subscript $c$, whenever they are not
explicitly needed. A contraction over color indices is implicit.
More general, the correlation functions contain a number of quark and gluon
fields, but at lowest order in $1/Q$ only the one above is relevant.

A point that needs to be mentioned is the color gauge invariance of the
correlation function defined above. For a proper gauge-invariant definition
of a nonlocal matrix element a color link operator,
$L(0,x)={\cal P}\,\exp [i g \int_0^x d s\cdot A(s)]$,
must be inserted. As will be seen below, we only need the integral of $\Phi$
over $k^-$, i.e., the nonlocality is restricted to the plane $x^+=0$.
Going further and integrating also over transverse
momenta, one is only sensitive to the nonlocality in a lightlike direction
(say $x^-$). With an appropriate choice of path (a straight link) and
lightcone gauge ($A^+= 0$), the link operator becomes just unity.
When transverse momentum is observed, this no longer is the case,
since one becomes sensitive to transverse separation $\bbox{x}_{T}$,
although still $x^+=0$. Fixing the residual gauge freedom that affects
$A_T$ can be achieved by imposing boundary conditions\cite{kogu70}.
A path from $0$ to $x$ can then be constructed such that the link operator
becomes unity after gauge-fixing.
How the choices of gauge and path connect to the proof
of factorization remains an open question.

\subsection{The Dirac structure of the correlation function}	\label{cfs}

In order to analyze the diagrams in Fig.~\ref{fig:DY} for DY scattering, we
need to investigate the Dirac structure of the correlation function. This can
be done by making an expansion in an appropriate basis. Constraints on the
correlation function come from hermiticity, parity invariance, and time
reversal invariance,
\begin{equation}\label{phicon}
\begin{array}{lcl}
\Phi^\dagger(P S;k)=\gamma^0\,\Phi(P S;k)\,\gamma^0&\qquad&
[\text{Hermiticity}]\\
\Phi(P S;k) = \gamma^0 \,\Phi(\bar P -\!\!\bar S;\bar k)\,\gamma^0 &&
[\text{Parity}] \\
\Phi^\ast(P S;k)=\gamma_5 C \,\Phi(\bar P\bar S;\bar k)\, C^\dagger\gamma_5&&
[\text{Time reversal}]
\end{array}
\end{equation}
where the charge conjugation matrix $C=i\gamma^2\gamma^0$,
and $\bar k^\mu=k_\mu$.
Choosing the Dirac matrix basis $\bf 1$, $i\gamma_5$, $\gamma^\mu$,
$\gamma^\mu\gamma_5$, and $i\sigma^{\mu\nu}\gamma_5$
(note that $\Gamma^\dagger=\gamma^0\Gamma\gamma^0$),
the most general structure satisfying these constraints is
\begin{eqnarray}
\Phi(P S;k) & = &
A_1\, \bbox{1} + A_2\, {\not\! P} + A_3\, {\not\! k}
\nonumber\\
&&+ A_4 \,\gamma_5 {\not\! S}
+ A_5\, \gamma_5 [{\not\! P},{\not\! S}]
+ A_6\, \gamma_5 [{\not\! k},{\not\! S}]\nonumber \\
& &
+ A_7\, k\cdot S\, \gamma_5  {\not\! P}
+ A_8\, k\cdot S\, \gamma_5 {\not\! k}
+ A_9\, k\cdot S\,\gamma_5  [{\not\! P},{\not\! k}].
\label{ampli}\end{eqnarray}
Hermiticity requires all the amplitudes $A_i=A_i(k\cdot P, k^2)$ to be
real. Note the presence of the amplitude $A_9$ which is left out in Eq.~(3.4)
of Ref.~\cite{rals79}.

The basic assumption
made for the correlation function is that in the hadron rest frame the
quark momentum $k$ is restricted to a hadronic scale $\Lambda$, explicitly
$k^2$ and $k\cdot P$ are of ${\cal O}(\Lambda^2)$.
In a frame where the hadron has no transverse momentum, the momentum $k$ is
written as
\begin{equation} \label{kmom}
k =\left[ \frac{k^2+\bbox{k}_{T}^2}{2x P^+}, x\,P^+, \bbox{k}_{T} \right],
\end{equation}
with the lightcone momentum fraction $x=k^+/P^+$. The restrictions on $k^2$
and $k\cdot P$ imply that also
$\bbox{k}_T^2=-k^2+2x k\cdot P-x^2 M^2$ is of ${\cal O}(\Lambda^2)$.
Considering diagram~\ref{fig:DY}$a$, one sees
easily that momentum conservation on the hard vertex implies $q^-=k_a^-+k_b^-$.
However, $k_b^-\sim P_B^-$, whereas $k_a^-\sim M_A^2/P_A^+$, which is
down by a factor $\sim M^2/Q^2$ in any collinear frame.
Therefore, for hadron $A$, one is led to study
$\int d k^-\,\Phi(P S;k)$, or equivalently its projections
$\int d k^-\,\text{Tr} [\Gamma \Phi]$. These latter quantities do not
carry Dirac indices anymore, but because of the $\Gamma$-matrices, they
{\em do\/} have a specific Lorentz tensor character.
Defining the projections
\begin{eqnarray}
&&\Phi[\Gamma](x,\bbox{k}_T)\equiv\frac{1}{2}\int d k^-\;\text{Tr}
\left[\Gamma\;\Phi\right]\\
&&\qquad\qquad=\frac{1}{2} \int \frac{d x^-}{2\pi}\frac{d^2
\bbox{x}_{T}}{(2\pi)^2}\;
\exp[i(xP^+ x^- -\bbox{k}_{T}\cdot\bbox{x}_{T})]\;\langle P S |
\overline{\psi}(0)\Gamma\psi(0,x^-,\bbox{x}_{T}) | P S\rangle  ,
\end{eqnarray}
one has for instance the vector projection
\begin{equation}
\Phi[\gamma^+]= \int d(2k\cdot P) dk^2\,
\delta\left(\bbox{k}_T^2 +k^2-2x k\cdot P+x^2 M^2\right)\,(A_2 + x\,A_3),
\end{equation}
which is ${\cal O}(1)$. Other projections, e.g., the scalar
\begin{equation}
\Phi[1]= \frac{1}{P^+}\int d(2k\cdot P) dk^2\,
\delta\left(\bbox{k}_T^2 +k^2-2x k\cdot P+x^2 M^2\right)\, A_1,
\end{equation}
contain an integral of ${\cal O}(1)$ multiplied by a factor $1/P^+$.
In the cross section, this factor will give rise to a suppression of order
$1/Q$.
In this way it is seen that the leading
contributions come from the Dirac structure where the number
of $+$-components minus the number of $-$-components is largest (that is, $1$).
They are parametrized as
\begin{eqnarray}
& & \Phi[\gamma^+]= f_1(x,\bbox{k}_{T}^2) ,\nonumber
\\ & & \Phi[\gamma^+ \gamma_5]=  g_{1L}(x,\bbox{k}_{T}^2) \lambda
+g_{1T}(x,\bbox{k}_{T}^2)\frac{\bbox{k}_T\cdot\bbox{S}_{T}}{M},
\label{tweetwist}
\\ & & \Phi[ i \sigma^{i+}\gamma_5 ]=h_{1T}(x,\bbox{k}_{T}^2) \bbox{S}_{T}^i
+ \left[h_{1L}^\perp(x,\bbox{k}_{T}^2)\lambda +h_{1T}^\perp(x,\bbox{k}_{T}^2)
\frac{\bbox{k}_T\cdot\bbox{S}_{T}}{M}\right]\frac{\bbox{k}_{T}^i}{M},\nonumber
\end{eqnarray}
defining six real {\em distribution functions\/}
per flavor, depending on $x$ and $\bbox{k}_{T}^2$.
These encode the leading behavior of the quark correlation function.
An expansion can be made in terms of local operators which have twist two and
higher.

In the diagrammatic expansion for the DY hadron tensor (with $A^+=0$ for the
lower blob), correlations will appear with, in addition to the quark fields,
transverse gluon fields in the soft part,
$\overline{\psi}(0)A_T^\alpha(y)\psi(x)$, etc.. They can be analyzed in the
same way, and turn out to be suppressed by one order of $1/Q$ for each
additional gluon field.

In summary, for a leading-order DY calculation, one is to use
\begin{equation}
\int d k^- \Phi = \case{1}{2} \Phi[\gamma^+]\;\gamma^-
+\case{1}{2}\Phi[\gamma^+\gamma_5] \;\gamma_5\gamma^-
+\case{1}{2}\Phi[i\sigma^{i+}\gamma_5] \;i\gamma_5\sigma^-\!_i+\ldots ,
\end{equation}
where $i$ is a transverse index (i.e., $i=1,2$), and the dots represent
projections that will come in only at ${\cal O}(1/Q)$.
At the leading-order level, one is left with only three
projections, which have vector, axial-vector, and axial-tensor character,
respectively, and which can be parametrized by six distribution functions.

\subsection{Antiquarks}

The antiquark correlation function, describing the antiquarks of flavor
$a$, is given by (contracting over color indices)
\begin{equation}\label{anticor}
(\overline{\Phi}_{\bar{a}/A})_{ij}(P_A S_A;k) =
\int \frac{d^4 x}{(2\pi)^4} \;\; e^{i k\cdot x}
\langle P_A S_A |  \psi_i^{(a)}(0)\overline{\psi}_j^{(a)}(x)|P_A S_A\rangle_c .
\end{equation}
Also the antiquark momentum $k$ can
be written as in Eq.~(\ref{kmom}). Its Dirac structure can be analyzed
likewise. We define the antiquark projections
\begin{eqnarray}
&&\overline{\Phi}[\Gamma](x,\bbox{k}_T)\equiv
\frac{1}{2} \int d k^- \; \text{Tr} \left[ \Gamma\; \overline{\Phi} \right]\\
&&\quad\qquad=\frac{1}{2} \int \frac{d x^-}{2\pi}\frac{d^2
\bbox{x}_{T}}{(2\pi)^2}\;
\exp[i(xP^+ x^- -\bbox{k}_{T}\cdot\bbox{x}_{T})]\;\langle P S |
\text{Tr} \left[\Gamma\psi(0) \overline{\psi}(0,x^-,\bbox{x}_{T})\right]
| P S\rangle.
\end{eqnarray}
Using the charge conjugation properties of Dirac fields and hadron states,
we deduce
\begin{equation}
\overline{\Phi}_{\bar{a}/A}=-C^{-1}\left(\Phi_{a/\bar{A}}\right)^T C.
\end{equation}
Upon demanding charge conjugation invariance of the distribution functions,
i.e., the quark distributions in the antihadron $\bar{A}$ are the same as the
corresponding antiquark distributions in $A$, we obtain the expressions
\begin{eqnarray}\label{antitwee}
& &\overline{\Phi}[\gamma^+]= \bar{f}_1(x,\bbox{k}_{T}^2) ,
\\& &\overline{\Phi}[\gamma^+ \gamma_5]= -\bar{g}_{1L}(x,\bbox{k}_{T}^2)
 \lambda-\bar{g}_{1T}(x,\bbox{k}_{T}^2)\frac{\bbox{k}_T\cdot\bbox{S}_{T}}{M},
\\ & &\overline{\Phi}[ i \sigma^{i+}\gamma_5 ]=
\bar{h}_{1T}(x,\bbox{k}_{T}^2) \bbox{S}_{T}^i+ \left[
\bar{h}_{1L}^\perp(x,\bbox{k}_{T}^2)\lambda +\bar{h}_{1T}^\perp
(x,\bbox{k}_{T}^2)\frac{\bbox{k}_T\cdot\bbox{S}_{T}}{M}\right]
\frac{\bbox{k}_{T}^i}{M}.
\end{eqnarray}

We note that the anticommutation
relations for fermions can be used to obtain the symmetry relation
\begin{equation}
\overline \Phi_{ij}(P S;k) = - \Phi_{ij}(P S;-k).
\end{equation}
For the distribution functions this gives the symmetry relation
\begin{equation}
\bar{f}_1(x,\bbox{k}_{T}^2) = - f_1(-x,\bbox{k}_{T}^2),
\end{equation}
and identically for $g_{1T}$, $h_{1T}$ and $h_{1T}^\perp$, whereas
\begin{equation}
\bar{g}_{1L}(x,\bbox{k}_{T}^2) = g_{1L}(-x,\bbox{k}_{T}^2),
\end{equation}
and identically for $h_{1L}^\perp$

Finally, we note that hadron $B$ can be treated in the same fashion.
However, since we have chosen to work in the collinear frames where
the third axis lies opposite to the direction of hadron $B$, the role of the
$+$- and $-$-components for $B$ must be interchanged as compared to $A$.
So, in Fig.~\ref{fig:DY}$a$ the upper blob reduces to $\int dk^+_b
\overline{\Phi}(P_B S_B;k_b)$, for which one has to use the gauge $A^-=0$.

\section{Free-quark ensemble}  			\label{Fqm}

The leading $\bbox{k}_{T}$-integrated distributions $f_1(x)$, $g_1(x)$,
and $h_1(x)$, have a parton model interpretation as the longitudinal momentum,
helicity, and transversity distribution, respectively.
In this section we show how, for a free-quark ensemble, this identification can
be generalized to the $\bbox{k}_{T}$-dependent distributions.

It is instructive to calculate the correlation function for a free-quark
target of flavor $a$. This is given by
\begin{equation}\label{quarkcor}
(\Phi_{a/a})_{ij}(p\; s;k) = \delta^4(k-p)\; u_i(p,s) \overline u_j(p,s)
= \delta^4(k-p) \left[ ({\not\! k} + m)\left(\frac{1 +
\gamma_5 {\not\! s}}{2}\right) \right]_{ij} ,
\end{equation}
where the momentum and spin vector are parametrized as
\begin{eqnarray}
k&=&\left[\frac{m^2 +\bbox{k}_{T}^2}{2k^+},k^+,\bbox{k}_{T}\right], \\
s & = & \lambda_a n_k + s_{at} =
\lambda_a\left[ \frac{\bbox{k}_{T}^2 - m^2}{2m\,k^+}, \frac{k^+}{m},
\frac{\bbox{k}_{T}}{m} \right]+\left[ \frac{\bbox{k}_{T}\cdot
\bbox{s}_{aT}}{k^+}, 0, \bbox{s}_{aT} \right].\label{paras}
\end{eqnarray}
We identify the lightcone helicity $\lambda_a$,
helicity vector $n_k$, and transverse polarization $s_{at}$
(transverse in the sense that $s_{at}\cdot k=s_{at}\cdot n_k=0$).
Note that the lightcone helicity vector $n_k$, satisfying $n_k\cdot k=0$ and
$n_k^2=-1$, acquires its conventional meaning~\cite[Eq.~(2-49)]{ItZub},
either if $\bbox{k}_{T}=\bbox{0}_{T}$, or in the infinite-momentum-limit
$k^+\rightarrow\infty$ with $\bbox{k}_{T}$ fixed.
One checks that $\lambda_a^2 + \bbox{s}_{aT}^2=-s^2 = 1$.
With the simple form~(\ref{quarkcor}) it is straightforward to calculate the
leading projections for a free-quark target with non-zero transverse
momentum,
\begin{eqnarray}
& & \frac{1}{2} \int dk^-\, \text{Tr}[\gamma^+ \Phi(p\; s;k)]
= \delta \left(\frac{k^+}{p^+}- 1 \right)\delta^2 (\bbox{k}_{T} -
\bbox{p}_{T}), \nonumber\\
& & \frac{1}{2} \int dk^-\,\text{Tr}[\gamma^+\gamma_5 \Phi(p\; s;k)]
= \delta \left(\frac{k^+}{p^+} - 1 \right)\delta^2 (\bbox{k}_{T} -
\bbox{p}_{T})\,\lambda_a, \\
& & \frac{1}{2} \int dk^-\,\text{Tr}[i\sigma^{i+}\gamma_5 \Phi(p\; s;k)]
=\delta \left(\frac{k^+}{p^+} - 1 \right)\delta^2 (\bbox{k}_{T} -
\bbox{p}_{T}) \,\bbox{s}_{aT}^i. \nonumber
\end{eqnarray}

This simple example of a quark target can be generalized to the case in
which the hadron is considered as a beam of non-interacting partons
of total momentum $P$ and angular momentum $S$. This is tantamount to
inserting free-field plane-wave expansions into the correlation
function~(\ref{corelgen}). One gets (summing over $\alpha,\beta=1,2$)
\begin{equation}\label{freeone}
\Phi_{ij}(P S;k)=2\,\delta(k^2-m^2) \left[\theta(k^+)
u^{(\beta)}_i(k){\cal P}_{\beta\alpha}(k)\bar{u}^{(\alpha)}_j(k)-\theta(-k^+)
v^{(\beta)}_i(-k)\overline{\cal P}_{\beta\alpha}(-k)\bar{v}^{(\alpha)}_j(-k)
\right].
\end{equation}
The functions ${\cal P}$ and $\overline{\cal P}$ are given by
\begin{eqnarray}\label{quarkden}
& & {\cal P}_{\beta\alpha}(k) = {\cal P}_{\beta \alpha}(x,\bbox{k}_{T})
\equiv\frac{1}{2(2\pi)^3} \int\frac{dx^\prime\,d^2\bbox{k}_{T}^\prime}
{(2\pi)^3\,2x^\prime}\,\langle PS|b^\dagger_{\alpha}(k^\prime)b_{\beta}(k)
|PS\rangle, \\
& & \overline{\cal P}_{\beta\alpha}(k)=\overline{\cal P}_{\beta \alpha}
(x,\bbox{k}_{T}) \equiv\frac{1}{2(2\pi)^3} \int
\frac{dx^\prime\,d^2\bbox{k}_{T}^\prime}{(2\pi)^3\,2x^\prime}\,
\langle PS|d^\dagger_{\beta}(k^\prime)d_{\alpha}(k)|PS\rangle.
\end{eqnarray}
The form of quantization one can choose to be instant-front, as well as
light-front quantization, since for free fields they are
equivalent~\cite{kogu70}. As for the choice of coordinates,
the use of lightcone coordinates is convenient, because of the integration
over $k^-$ that is needed in deep inelastic processes.
The Dirac structure can be parametrized as
\begin{eqnarray}
& & u^{(\beta)}(k) {\cal P}_{\beta \alpha}(k) \bar u^{(\alpha)}(k)
= {\cal P}(k) ({\not\! k} + m) \left( \frac{1 + \gamma_5 {\not\! s} (k)}{2}
\right), \\
& & v^{(\beta)}(k) \overline{\cal P}_{\beta \alpha}(k) \bar v^{(\alpha)}(k)
= \overline{\cal P}(k) ({\not\! k} - m) \left( \frac{1 + \gamma_5
\overline{\not\! s}(k)}{2}\right),
\end{eqnarray}
in terms of  positive definite quark and antiquark probability densities
${\cal P}(k)$ and $\overline{\cal P}(k)$, and spin vectors $s^\mu(k)$ and
$\overline s^\mu(k)$ . Inserting the free-field expansion in the current
expectation value $\langle PS|\overline{\psi}(0)\gamma^\mu\psi(0)|PS\rangle
=2P^\mu (N -\overline{N})$, where $N$ and $\overline{N}$ are the total number
of quarks and antiquarks, respectively, one obtains from the $+$-component
the normalizations
$\int_0^1 dx \int d^2\bbox{k}_{T} {\cal P}(x,\bbox{k}_{T}^2) = N$ and
$\int_0^1 dx \int d^2\bbox{k}_{T}\overline{\cal P}(x,\bbox{k}_{T}^2)
=\overline{N}$.
The average quark spin vector $s^\mu(k)$ is parametrized by the
helicity density $\lambda_a(x,\bbox{k}_{T})$
and transverse polarization density
$\bbox{s}_{aT}(x,\bbox{k}_{T})$ by expanding $s^\mu$ as in Eq.~(\ref{paras}).
A similar parametrization is used for $\overline{s}^\mu(k)$ in terms of
$\lambda_{\bar{a}}(x,\bbox{k}_{T})$ and $\bbox{s}_{\bar{a}T}(x,\bbox{k}_{T})$.

Integrating Eq.~(\ref{freeone}) over $k^-$ one obtains the result for a
free-quark ensemble,
\begin{eqnarray}
\frac{1}{2}\int dk^- \Phi(k)&=&\theta(x)\frac{{\cal P}(x,\bbox{k}_{T}^2)}{2k^+}
({\not\! k}+m)\left(\frac{1+\gamma_5 {\not\! s}(x,\bbox{k}_{T})}{2}\right)
\nonumber\\
&-&\theta(-x)\frac{\overline{\cal P}(-x,\bbox{k}_{T}^2)}{2k^+}({\not\! k}+m)
\left(\frac{1+\gamma_5 \overline{\not\! s}(-x,-\bbox{k}_{T})}{2}\right).
\label{lcfree}
\end{eqnarray}
This gives (for $x > 0$)
\begin{eqnarray}
& & \frac{1}{2} \int dk^-\,\text{Tr}[\gamma^+ \Phi(P S;k)]
= {\cal P}(x,\bbox{k}_{T}^2), \nonumber\\
& & \frac{1}{2} \int dk^-\,\text{Tr}[\gamma^+\gamma_5 \Phi(P S;k)]
=  {\cal P}(x,\bbox{k}_{T}^2)\, \lambda_a(x,\bbox{k}_{T}), \\
& & \frac{1}{2} \int dk^-\,\text{Tr}[i\sigma^{i+}\gamma_5 \Phi(P S;k)]
= {\cal P}(x,\bbox{k}_{T}^2)\,\bbox{s}_{aT}^i(x,\bbox{k}_{T})  .\nonumber
\end{eqnarray}
A comparison with Eq.~(\ref{tweetwist}) yields
\begin{eqnarray}
& & {\cal P}(x,\bbox{k}_{T}^2) = f(x,\bbox{k}_{T}^2), \nonumber\\
& & {\cal P}(x,\bbox{k}_{T}^2)\, \lambda_a(x,\bbox{k}_{T}) =
g_{1L}(x,\bbox{k}_{T}^2)\,\lambda
+  g_{1T}(x,\bbox{k}_{T}^2)\,\frac{\bbox{k}_T\cdot\bbox{S}_{T}}{M}\,,\\
& &{\cal P}(x,\bbox{k}_{T}^2) \, \bbox{s}_{aT}^i(x,\bbox{k}_{T}) \,
= h_{1T}(x,\bbox{k}_{T}^2)\,\bbox{S}_{T}^i\, + \left[
h_{1L}^\perp(x,\bbox{k}_{T}^2)\lambda +h_{1T}^\perp(x,\bbox{k}_{T}^2)
\frac{\bbox{k}_T\cdot\bbox{S}_{T}}{M}\right]\,\frac{\bbox{k}_{T}^i}{M},
\nonumber
\end{eqnarray}
which shows how for $x>0$ the functions $g_{1L}$, $g_{1T}$, $h_{1T}$,
$h_{1L}^\perp$, and $h_{1T}^\perp$, are to be interpreted as quark
longitudinal and transverse polarization distributions. The function
$h_{1T}^\perp$ was omitted in the paper of RS. Specifically for non-zero
transverse momenta of the quarks, it becomes relevant.
For the antiquarks the same relations hold between the antiquark probability
density $\overline{\cal P}$, helicity density $\lambda_{\bar{a}}$,
and transverse polarization density $\bbox{s}_{\bar{a}T}$, on the one hand,
and the antiquark distributions on the other hand.
Extending to all $x$, results are obtained in accordance with the symmetry
relations in the previous section, e.g.,
$f(x,\bbox{k}_{T}^2)=\theta(x) {\cal P}(x,\bbox{k}_{T}^2)
- \theta(-x)\overline{\cal P} (-x,\bbox{k}_{T}^2)$.
For on-shell quarks, relations exist between the distributions due to
constraints from Lorentz invariance~\cite{EFP}. The scalar $\cal P$ can
only depend on $2k\cdot P=m^2/x+M^2 x+\bbox{k}_T^2/x$ and the pseudovector
$s^\mu$ can be parametrized by two functions depending on the same combination
of $x$ and $\bbox{k}_T^2$.

We note that for the leading-order matrix elements the free-field results
can be used to provide a parton interpretation, even in the interacting theory,
because the distribution functions can be expressed as densities for
specific projections of the so-called `good' components of the quark field;
$\psi_+ \equiv \Lambda_+ \psi$, where $\Lambda_+=\case{1}{2} \gamma^-\gamma^+
$.
In light-front quantization a Fourier expansion for the good components
(at $x^+= 0$) can be written down in which the Fourier coefficients can be
interpreted as particle and antiparticle creation and annihilation
operators~\cite{kogu70}. The different polarization distributions involve
projection operators that commute with $\Lambda_+$~\cite{jaji}.

At subleading order,
the analysis of the quark-quark correlation functions leads to a number
of new distribution functions. For free quarks, they can
also be expressed in the quark densities and thus they can be related to
the leading distribution functions. However, it turns out that
the presence of nonvanishing quark-quark-gluon correlation
functions causes deviations from the free-field results~\cite{tang94}.

\section{Results}					\label{Res}

\subsection{Hadron tensor}				\label{had}

Using the EFP-expansion, the leading-order Drell-Yan hadron tensor
in the deep inelastic limit, and with $Q_T={\cal O}(\Lambda)$, can
be written as the sum of the quark and antiquark Born diagrams in
Fig.~\ref{fig:DY}. First, we will calculate diagram~\ref{fig:DY}$a$,
in which a quark of hadron $A$ annihilates an antiquark of $B$.
It reads
\begin{equation}\label{Born}
W_{\text{quark}}^{\mu\nu} =\frac{1}{3}\sum_{a,b}  \delta_{b\bar{a}}e^2_a
\int d^4k_a\, d^4k_b\;\delta^4(k_a + k_b -q)
\;\text{Tr} \left[ \Phi_{a/A} (P_A S_A; k_a)\; \gamma^\mu \;
\overline{\Phi}_{b/B}(P_B S_B;k_b)\; \gamma^\nu \right] ,
\end{equation}
where $a$ ($b$) runs over all quark (antiquark) flavors, and $e_a$ is the
quark charge in units of $e$. The factor ${1}/{3}$ comes from the fact that
the quark fields in both the correlation functions are traced over a color
identity operator, which is appropriate since only color-singlet operators
can give non-zero matrix elements between (color-singlet) hadron states
(Wigner-Eckart theorem).  However, since the diagrams we consider have
only one quark loop, one has only one color summation, leading to a color
factor ${1}/{3}$.
Using the boundedness of quark momenta in hadrons as discussed before,
one  finds $k_a^+ \gg k_b^+$ and $k_b^- \gg k_a^-$. Thus, the delta function
can be approximated by
\begin{equation}\label{delta}
\delta^4(k_a + k_b -q)\approx\delta(k_a^+-q^+)\delta(k_b^- -q^-)\delta^2
(\bbox{k}_{aT}+\bbox{k}_{bT}-\bbox{q}_{T}).
\end{equation}
It follows that indeed we are sensitive to $\int dk^- \Phi_A(k)$ and
$\int dk^+ \overline{\Phi}_B(k)$. Furthermore, the trace in Eq.~(\ref{Born})
can be factorized by means of the Fierz decomposition
\begin{eqnarray}
\lefteqn{4(\gamma^\mu)_{jk}(\gamma^\nu)_{li}=}\nonumber\\
&&\left[ {\bf 1}_{ji}{\bf 1}_{lk}
+(i\gamma_5)_{ji}(i\gamma_5)_{lk}-(\gamma^\alpha)_{ji}(\gamma_\alpha)_{lk}
-(\gamma^\alpha\gamma_5)_{ji}(\gamma_\alpha\gamma_5)_{lk}
+\case{1}{2} (i\sigma_{\alpha\beta}\gamma_5)_{ji}(i\sigma^{\alpha\beta}
\gamma_5)_{lk}\right]g^{\mu\nu} \nonumber\\
&&+(\gamma^{\{ \mu})_{ji}(\gamma^{\nu\} })_{lk}+
(\gamma^{\{ \mu}\gamma_5)_{ji}(\gamma^{\nu\} }\gamma_5)_{lk}+
(i\sigma^{\alpha\{\mu}\gamma_5)_{ji}(i{\sigma^{\nu\}} }_\alpha \gamma_5)_{lk}
+\ldots,\label{Dirac}
\end{eqnarray}
where the dots denote structures antisymmetric under the exchange of
$\mu$ and $\nu$. We keep only the leading projections, as they where found
in Sec.~\ref{For}. This leads to
\begin{eqnarray}\label{compact}
W_{\text{quark}}^{\mu\nu} &=&-\frac{1}{3}\sum_{a,b}  \delta_{b\bar{a}}e^2_a
\int d^2\bbox{k}_{aT}\, d^2\bbox{k}_{bT}\; \delta^2(\bbox{k}_{aT}
+\bbox{k}_{bT}-\bbox{q}_{T})\nonumber\\
&&\biggl( \Phi_{a/A}[\gamma^+]\,\overline{\Phi}_{b/B}[\gamma^-]+
\Phi_{a/A}[\gamma^+\gamma_5]\,\overline{\Phi}_{b/B}[\gamma^-\gamma_5]
\biggr)g_T^{\mu\nu}\nonumber\\
&&+\Phi_{a/A}[i\sigma^{i+}\gamma_5]\,\overline{\Phi}_{b/B}[i\sigma^{j-}
\gamma_5]\left(g_{T i}\!^{\,\{\mu}g_{T}\!^{\nu\}}\!_j
-g_{T ij}g_T^{\mu\nu}\right)+{\cal O}\left(\frac{1}{Q}\right),
\end{eqnarray}
where the $\Phi_{a/A}$-projections are taken at $x=x_A$ and $\bbox{k}_{aT}$,
the $\overline{\Phi}_{b/B}$-projections at $x=x_B$ and $\bbox{k}_{bT}$.
The next step would be to insert Eq.~(\ref{tweetwist}) and their hadron-$B$
antiquark counterparts. In this leading-order calculation we may interchange
the subscript $T$ by a $\perp$ where we wish, because of
Eqs.~(\ref{Pmetric}),~(\ref{Pvector}), and~(\ref{Pinprod}).
The resulting expression contains convolutions like
\begin{equation}\label{example}
\int d^2\bbox{k}_{aT}\,d^2\bbox{k}_{bT}\;\delta^2(\bbox{k}_{aT}+\bbox{k}_{bT}
-\bbox{q}_{T})F(\bbox{k}_{aT}^2,\bbox{k}_{bT}^2)k_{a\perp}^\mu.
\end{equation}
They are not of the desired form, since the (perpendicular) Lorentz index is
carried by a convolution variable.
In order to write the hadron tensor in terms of structure functions,
the Lorentz indices must be carried by external vectors, like $X^\mu$.
In the appendix we describe the method we used to project perpendicular
Lorentz tensors, like~(\ref{example}), onto the $XY$-basis.
For instance, (\ref{example}) can be written as (up to order $1/Q^2$)
\begin{equation}
\hat{x}^\mu \int d^2\bbox{k}_{aT}\,d^2\bbox{k}_{bT}\;\delta^2(\bbox{k}_{aT}
+\bbox{k}_{bT}-\bbox{q}_{T})F(\bbox{k}_{aT}^2,\bbox{k}_{bT}^2)
\left(\frac{Q_T^2+\bbox{k}_{aT}^2-\bbox{k}_{bT}^2}{2Q_T}\right).
\end{equation}

Before presenting the results, we mention that the antiquark
diagram~\ref{fig:DY}$b$ can be obtained from the quark diagram~\ref{fig:DY}$a$
by making the replacements\footnote{It is seen from Eq.~(\ref{compact})
that this amounts to replacing $A\leftrightarrow B$, rendering a result,
symmetric under the exchange of the two hadrons.}
$a\leftrightarrow \bar{a}$, $f\leftrightarrow\bar{f}$,
$g\leftrightarrow -\bar{g}$, $h\leftrightarrow \bar{h}$,
where $g$ is generic for the quark axial-vector distributions, etcetera.
It is convenient to define the convolution of two arbitrary distributions
$d_1(x,\bbox{k}_{T}^2)$ and $\bar{d}_2(x,\bbox{k}_{T}^2)$ (possibly
multiplied by an overall function of $\bbox{k}_{aT}^2$ and $\bbox{k}_{bT}^2$)
\begin{equation}\label{Iconv}
I[d_1\bar{d}_2]\equiv\frac{1}{3}\sum_{a,b} \delta_{b\bar{a}}e^2_a \int d^2
\bbox{k}_{aT}\, d^2\bbox{k}_{bT}\;\delta^2(\bbox{k}_{aT}+\bbox{k}_{bT}
-\bbox{q}_{T})d_1(x_A,\bbox{k}_{aT}^2)\bar{d}_2(x_B,\bbox{k}_{bT}^2),
\end{equation}
summing over quark {\em and\/} antiquark flavors, with the prescription
that if its argument is of antiquark (quark) nature, then $d_1$ ($\bar{d}_2$)
is to be replaced, either by $\bar{d}_1$ ($d_2$) if it concerns a vector or
axial-tensor distribution, or by $-\bar{d}_1$ ($-d_2$) if it concerns an
axial-vector distribution.
We split up the general leading-order hadron tensor $W^{\mu\nu}(S_A,S_B)$ in
four cases, corresponding to unpolarized, longitudinal-longitudinal,
transverse-longitudinal, and transverse-transverse scattering,
\begin{eqnarray}
W^{\mu\nu}(0,0)&=&-W_T\, g_\perp^{\mu\nu},\label{WUU}\\
W^{\mu\nu}(\lambda_A,\lambda_B)&=&-\left(W_T+\case{1}{4}V^{LL}_T
\lambda_A\lambda_B\right)g_\perp^{\mu\nu}
-\case{1}{4}V^{LL}_{2,2}\lambda_A\lambda_B
(\hat{x}^\mu\hat{x}^\nu+\case{1}{2}g_\perp^{\mu\nu}),\\
W^{\mu\nu}(\bbox{S}_{AT},\lambda_B)&=&
-\left[W_T -V^{TL}_T(\hat{x}\cdot S_{A\perp})\lambda_B
\right]g_\perp^{\mu\nu}
+V^{TL}_{2,2}(\hat{x}\cdot S_{A\perp})\lambda_B(\hat{x}^\mu\hat{x}^\nu
+\case{1}{2}g_\perp^{\mu\nu})\nonumber\\
&&- U^{TL}_{2,2}\lambda_B\left[\hat{x}^{\{\mu}S_{A\perp}^{\nu\}}
-(\hat{x}\cdot S_{A\perp})g_\perp^{\mu\nu}\right],\\
W^{\mu\nu}(\bbox{S}_{AT},\bbox{S}_{BT})&=&
-\left[W_T - V^{1\,TT}_T (S_{A\perp}\cdot S_{B\perp})
+V^{2\,TT}_T (\hat{x}\cdot S_{A\perp})(\hat{x}\cdot S_{B\perp})\right]
g_\perp^{\mu\nu}\nonumber\\
&&+\left[V^{1\, TT}_{2,2}(S_{A\perp}\cdot S_{B\perp})
- V^{2\,TT}_{2,2}(\hat{x}\cdot S_{A\perp})(\hat{x}\cdot S_{B\perp})\right]
(\hat{x}^\mu\hat{x}^\nu+\case{1}{2} g_\perp^{\mu\nu})\nonumber\\
&&+U^{A\, TT}_{2,2}(\hat{x}\cdot S_{B\perp})
\left[\hat{x}^{\{\mu}S_{A\perp}^{\nu\}}
-(\hat{x}\cdot S_{A\perp})g_\perp^{\mu\nu}\right]\nonumber\\
&&+U^{B\, TT}_{2,2}(\hat{x}\cdot S_{A\perp})
\left[\hat{x}^{\{\mu}S_{B\perp}^{\nu\}}
-(\hat{x}\cdot S_{B\perp})g_\perp^{\mu\nu}\right]\nonumber\\
&&-U^{TT}_{2,2}\left[S_{A\perp}^{\{\mu}S_{B\perp}^{\nu\}}-(S_{A\perp}\cdot
S_{B\perp})g_\perp^{\mu\nu}\right],  \label{WTT}
\end{eqnarray}
where the structure functions, depending on $x_A$, $x_B$, and $Q_T$,
are given by
\begin{mathletters}\label{mies}
\begin{eqnarray}
& & W_T = I[f_1 \bar{f}_1],\label{uu}\\
& & V^{LL}_T = -4I[g_{1L}\bar{g}_{1L}],\label{Tll}\\
& & V^{LL}_{2,2} = I[\left(\alpha +\beta -\frac{(\alpha-\beta)^2}
{Q_T^2}\right)\frac{4 h_{1L}^\perp \bar{h}_{1L}^\perp}{M_A M_B} ],\label{Dll}\\
& & V^{TL}_T = I[\left(-Q_T^2 -\alpha+\beta
\right)\frac{g_{1T}\bar{g}_{1L}}{2M_A Q_T}],\\
& & V^{TL}_{2,2} = I[\left(\beta Q_T^2 +\alpha^2+\alpha\beta
-2\beta^2-\frac{(\alpha-\beta)^3}{Q_T^2}
\right)\frac{h_{1T}^\perp \bar{h}_{1L}^\perp}{M_A^2M_B Q_T}],\\
& & U^{TL}_{2,2} = I[(Q_T^2 -\alpha+\beta )\frac{h_{1T} \bar{h}_{1L}^\perp}
{2M_B Q_T} \nonumber\\&&\qquad\qquad
+\left(Q_T^2 (\alpha -\beta)-2(\alpha^2 -\beta^2)
+\frac{(\alpha-\beta)^3}{Q_T^2}\right)\frac{h_{1T}^\perp
\bar{h}_{1L}^\perp}{4M_A^2 M_B Q_T}],\\
& & V^{1\, TT}_T = I[\left(-Q_T^2 +2\alpha + 2\beta -\frac{(\alpha-\beta)^2}
{Q_T^2}\right)\frac{g_{1T}\bar{g}_{1T}}{4M_A M_B}],\\
& & V^{2\, TT}_T = I[\left(-\alpha-\beta+\frac{(\alpha-\beta)^2}{Q_T^2}\right)
\frac{g_{1T}\bar{g}_{1T}}{2M_A M_B}],\\
& & V^{2\, TT}_{2,2} = I[\left(\alpha^2+\beta^2 -\frac{2(\alpha+\beta)
(\alpha-\beta)^2}{Q_T^2}+\frac{(\alpha-\beta)^4}{Q_T^4}\right)
\frac{h_{1T}^\perp\bar{h}_{1T}^\perp}{M_A^2 M_B^2}],\\
& & V^{1\, TT}_{2,2}+ U^{A\, TT}_{2,2}+U^{B\, TT}_{2,2}=
\nonumber\\&&\qquad\qquad
I[\left( Q_T^2-2\alpha +\frac{(\alpha-\beta)^2}{Q_T^2}\right)\frac{h_{1T}
\bar{h}_{1T}^\perp}{2M_B^2}+\left( Q_T^2-2\beta
+\frac{(\alpha-\beta)^2}{Q_T^2}
\right)\frac{h_{1T}^\perp \bar{h}_{1T}}{2M_A^2}\nonumber\\
&&\qquad\qquad +\left(Q_T^2 (\alpha +\beta) -4\alpha^2 -4\beta^2 +
\frac{5(\alpha+\beta)
(\alpha-\beta)^2}{Q_T^2}-\frac{2(\alpha-\beta)^4}{Q_T^4}\right)
\frac{h_{1T}^\perp \bar{h}_{1T}^\perp}{4M_A^2 M_B^2}],\\
& & U^{A\, TT}_{2,2}-U^{B\, TT}_{2,2} =
I[\left( Q_T^2-2\alpha +\frac{(\alpha-\beta)^2}{Q_T^2}\right)\frac{h_{1T}
\bar{h}_{1T}^\perp}{2M_B^2}-\left( Q_T^2-2\beta
+\frac{(\alpha-\beta)^2}{Q_T^2}
\right)\frac{h_{1T}^\perp \bar{h}_{1T}}{2M_A^2}\nonumber\\
&&\qquad\qquad +\left( Q_T^2 (\alpha -\beta) -2\alpha^2 +2\beta^2
+\frac{(\alpha-\beta)^3}
{Q_T^2}\right)\frac{h_{1T}^\perp\bar{h}_{1T}^\perp}{4M_A^2 M_B^2}],\\
& & U^{TT}_{2,2}+\case{1}{2}U^{A\, TT}_{2,2}+\case{1}{2}U^{B\, TT}_{2,2}=
I[h_{1T} \bar{h}_{1T}+\alpha\frac{h_{1T}^\perp\bar{h}_{1T}}{2M_A^2}
+\beta\frac{h_{1T}\bar{h}_{1T}^\perp}{2M_B^2}
\nonumber\\&&\qquad\qquad
-\left( (\alpha-\beta)^2-\,\frac{2(\alpha+\beta)
(\alpha-\beta)^2}{Q_T^2}+\frac{(\alpha-\beta)^4}{Q_T^4}\right)
\frac{h_{1T}^\perp\bar{h}_{1T}^\perp}{8M_A^2 M_B^2}],
\end{eqnarray}
\end{mathletters}
where $\alpha =\bbox{k}_{aT}^2$, and $\beta=\bbox{k}_{bT}^2$.
In the last three equations we only gave the combination of structure
functions that will show up independently in the cross section,
to be discussed in the next subsection.
As for the nomenclature, we based ourselves on the RS-notation, except that we
{\em always\/} denote both hadron polarizations by the $TL$-superindices.
So, for instance, their $V_{0,0}^L$ corresponds to our $V_{0,0}^{LL}$,
and their $U_{2,2}$ equals our $U_{2,2}^{TT}$. The subindex $T$ stands for
the linear combination $(0,0)-\case{1}{3} (2,0)$.
Note that the hadron tensors~(\ref{WUU})-(\ref{WTT}) manifestly
conserve the electromagnetic current, i.e., $q_\mu W^{\mu\nu}=0$.

\subsection{Cross sections}				\label{cro}

It is now a straightforward matter to contract the hadron tensor
with the angle-dependent leptonic tensor~(\ref{hoeklep}). For the four
combinations of polarization we deduce the following leading-order
differential cross sections:
\begin{eqnarray}
\frac{d\sigma (0,0)}{d^4 q d\Omega}=\frac{\alpha^2}{2sQ^2}&&
W_T (1+\cos^2\theta),\label{unun}\\
\frac{d\sigma (\lambda_A,\lambda_B)}{d^4 q d\Omega}=
\frac{\alpha^2}{2sQ^2}&&\Bigl[
\left( W_T +\case{1}{4}V^{LL}_T\lambda_A\lambda_B
\right)(1+\cos^2\theta)
+ \case{1}{8}V^{LL}_{2,2}\lambda_A\lambda_B\sin^2\theta\cos 2\phi
\Bigr],\label{lolo}\\
\frac{d\sigma (\bbox{S}_{AT},\lambda_B)}{d^4 q d\Omega}=\frac{\alpha^2}{2sQ^2}
&&\Bigl[\left(W_T+V^{TL}_T\lambda_B \cos\phi_A
\right)(1+\cos^2\theta)+ \case{1}{2}V^{TL}_{2,2}
\lambda_B\cos\phi_A\sin^2\theta\cos 2\phi\nonumber\\
&& + U^{TL}_{2,2}\lambda_B\sin^2\theta\cos(2\phi-\phi_A)\Bigr],\\
\frac{d\sigma (\bbox{S}_{AT},\bbox{S}_{BT})}{d^4 q d\Omega}
=\frac{\alpha^2}{2sQ^2}
&&\Bigl[ \Bigl(W_T+ V^{1\, TT}_T \cos (\phi_A-\phi_B)
+V^{2\, TT}_T \cos\phi_A\cos\phi_B\Bigr)(1+\cos^2\theta)\nonumber\\
&&+\case{1}{2}V^{2\, TT}_{2,2}\cos\phi_A\cos\phi_B \sin^2\theta\cos 2\phi
\nonumber\\
&& +\case{1}{2}(V^{1\, TT}_{2,2}+ U^{A\, TT}_{2,2}+U^{B\, TT}_{2,2})
\cos(\phi_A-\phi_B)\sin^2\theta\cos 2\phi\nonumber\\
&& +\case{1}{2} (U^{A\, TT}_{2,2}-U^{B\, TT}_{2,2}) \sin(\phi_A - \phi_B)\,
\sin^2\theta\sin 2\phi\nonumber\\
&& +(U^{TT}_{2,2}+\case{1}{2}U^{A\, TT}_{2,2}+\case{1}{2}U^{B\, TT}_{2,2})
\sin^2\theta\cos (2\phi-\phi_A-\phi_B) \Bigr],\label{trtr}
\end{eqnarray}
where $\phi_A$ ($\phi_B$) is the azimuthal angle of $S_{A\perp}$
($S_{B\perp}$),
i.e., $\cos \phi_A=-\hat{x}\cdot S_{A\perp}\approx \bbox{q}_{T}\cdot
\bbox{S}_{AT}/Q_T$ for pure transverse polarization ($|\bbox{S}_{AT}|=1$).
Eq.~(\ref{trtr}) shows that the structure functions $V^{1\, TT}_{2,2}$,
$U^{A\, TT}_{2,2}$, $U^{B\, TT}_{2,2}$, and $U^{TT}_{2,2}$ cannot separately be
extracted from the experiment,
but only in three particular combinations. This is due to
the fact that the corresponding tensor structures in Eq.~(\ref{WTT}) are not
independent, because of the relation
\begin{equation}
g_\perp^{\rho\sigma}(2\hat{x}^\mu\hat{x}^\nu+g_\perp^{\mu\nu})
-\hat{x}^\rho(\hat{x}^{\{\mu}g_\perp^{\nu\}\sigma}-\hat{x}^\sigma
g_\perp^{\mu\nu})
-\hat{x}^\sigma(\hat{x}^{\{\mu}g_\perp^{\nu\}\rho}-\hat{x}^\rho
g_\perp^{\mu\nu})
-(g_\perp^{\rho\{\mu}g_\perp^{\nu\}\sigma}-g_\perp^{\rho\sigma}
g_\perp^{\mu\nu})=0,
\end{equation}
which can be proven by using $g_\perp^{\mu\nu}=-\hat{x}^\mu\hat{x}^\nu
-\hat{y}^\mu\hat{y}^\nu$. One easily checks that the angular functions of
Eqs.~(\ref{unun})-(\ref{trtr}) have no further dependencies.

In order to circumvent normalization problems, in spin experiments
one usually considers the asymmetries
\begin{equation}
A_{S_A S_B}=\frac{\sigma(S_A,S_B)-\sigma(S_A,-S_B)}{\sigma(S_A,S_B)
+\sigma(S_A,-S_B)}.
\end{equation}
For example, the longitudinal-longitudinal asymmetry follows from
Eq.~(\ref{lolo}),
\begin{equation}
A_{\lambda_A\lambda_B}=\frac{1}{4}\lambda_A\lambda_B
\left[\frac{V^{LL}_T}{W_T}
+\frac{\sin^2\theta\cos2\phi}{1+\cos^2\theta}\frac{\case{1}{2}V^{LL}_{2,2}}
{W_T}\right],\label{llas}
\end{equation}
where the structure functions can be found in Eqs.~(\ref{uu}),~(\ref{Tll}),
and~(\ref{Dll}). The angle-independent term is a generalization of the
$\bbox{q}_{T}$-integrated longitudinal-longitudinal asymmetry~\cite{clos77}.

\section{Discussion and conclusion}				\label{Dis}

In this last section we will look more closely into the results we derived,
first by comparing the cross sections with those integrated over
$\bbox{q}_{T}$, secondly by considering the limit $Q_T\rightarrow 0$,
and finally by considering a Gaussian $\bbox{k}_{T}^2$-dependence of the
distribution functions.

\subsection{$Q_T$-integrated results}

If the $Q_T$-dependence is eliminated by integrating over the transverse
momentum of the produced lepton pair, one recovers the lightcone momentum,
helicity, and transversity distributions of Eqs.~(\ref{feenx}),
(\ref{geenx}), and~(\ref{heenx});
\begin{eqnarray}
f_1(x)&=&\int d^2\bbox{k}_{T}\; f_1(x,\bbox{k}_{T}^2),\\
g_1(x)&=&\int d^2\bbox{k}_{T}\; g_{1L}(x,\bbox{k}_{T}^2),\\
h_1(x)&=&\int d^2\bbox{k}_{T} \left[h_{1T}(x,\bbox{k}_{T}^2)
+\frac{\bbox{k}_{T}^2}{2M^2}h_{1T}^\perp(x,\bbox{k}_{T}^2)\right],
\end{eqnarray}
and similar antiquark distributions. This can be seen in two ways.
It is easiest to return to Eq.~(\ref{compact}) and observe that the
delta function is absorbed by the $\bbox{q}_{T}$-integration,
so that the transverse
intergrations over $\bbox{k}_{aT}$ and $\bbox{k}_{bT}$ separate.
One is left with three independent structure functions,
\begin{mathletters}\label{intsf}
\begin{eqnarray}
\overline{W}_T(x_A,x_B)&=&\case{1}{3}\sum_a e_a^2\; f_1(x_A)
\bar{f}_1(x_B),\\
\overline{V}_T^{LL}(x_A,x_B)&=&-\case{4}{3}\sum_a e_a^2\; g_1(x_A)
\bar{g}_1(x_B),\\
\overline{U}_{2,2}^{TT}(x_A,x_B)&=&\case{1}{3}\sum_a e_a^2 \; h_1(x_A)
\bar{h}_1(x_B),
\end{eqnarray}
\end{mathletters}
with the by now well-known prescription that if $a$ is an antiflavor, the
vector and axial-tensor distributions have to be replaced by their charge
conjugated partners, the axial-vector distributions by minus their
anti-partners. The names of the structure functions in Eq.~(\ref{intsf}) refer
to the Lorentz tensor structure they multiply in the hadron tensor (and hence
the angular distribution in the cross section). That is, $\overline W_T$
in the hadron tensor is the coefficient of $-g_\perp^{\mu\nu}$, etcetera.
The same expressions are obtained if one integrates the full results in
Eqs.~(\ref{WUU})-(\ref{WTT}) over $\bbox{q}_{T}$. It is then seen that
$\overline{W}_T$ and $\overline{V}_T^{LL}$ are the integrals of $W_T$
and $V_T^{LL}$, respectively, but that $\overline{U}^{TT}_{2,2}$ originates
from
the combination $U^{TT}_{2,2} + \case{1}{2} U^{A\, TT}_{2,2}
+\case{1}{2} U^{B\, TT}_{2,2}+\case{1}{8} V^{2\, TT}_{2,2}$.

Contraction of the hadron tensors with the lepton tensor leads to the
following leading-order four-fold differential cross sections~\cite{rals79}:
\begin{eqnarray}
\frac{d\sigma (0,0)}{d x_A d x_B d\Omega}&=&\frac{\alpha^2}{4Q^2}
\overline W_T(1+\cos^2\theta),\\
\frac{d\sigma (\lambda_A,\lambda_B)}{d x_A d x_B d\Omega}
&=&\frac{\alpha^2}{4Q^2}
(\overline W_T+\case{1}{4}\overline V_T^{LL}\lambda_A\lambda_B)
(1+\cos^2\theta),\\
\frac{d\sigma (\bbox{S}_{AT},\lambda_B)}{d x_A d x_B d\Omega}
&=&\frac{\alpha^2}{4Q^2}\overline W_T(1+\cos^2\theta) ,\\
\frac{d\sigma (\bbox{S}_{AT},\bbox{S}_{BT})}{dx_A d x_B d\Omega}
&=&\frac{\alpha^2}{4Q^2}\left[\overline W_T(1+\cos^2\theta)
+ \overline U^{TT}_{2,2}\sin^2\theta\cos (2\phi -\phi_A-\phi_B)\right].
\end{eqnarray}
Note that no absolute azimuthal angles occur, but only the relative angles
$\phi -\phi_A$ and $\phi -\phi_B$. Two remarks are in place here.
Comparing with the full cross sections~(\ref{unun}-\ref{trtr}),
we observe the disappearance of the $\sin^2\theta\cos 2\phi$-term in
the longitudinal-longitudinal cross section, and hence in the
asymmetry~(\ref{llas}).
Secondly, we see no polarization-dependence in the integrated
transverse-longitudinal cross section, and hence no
asymmetry. Indeed this asymmetry is suppressed by a factor
$\sim 1/Q$~\cite{jaji}.

\subsection{$Q_T=0$}

Only a few of the structure functions are nonvanishing at $Q_T = 0$,
the other ones having kinematical zeros (see also~\cite{lam78}).
Therefore, it is worthwhile to consider the limit $Q_T \rightarrow 0$
of the structure functions in Eq.~(\ref{mies}).
In general we can do this, rewriting the convolutions in Eq.~(\ref{Iconv})
by making a transformation to the momenta $\bbox{k}_{T}=\case{1}{2}
(\bbox{k}_{aT}-\bbox{k}_{bT})$ and $\bbox{K}_T=\bbox{k}_{aT}+\bbox{k}_{bT}$,
\begin{eqnarray}
I[d_1\bar{d}_2]&=&\case{1}{3}\sum_{a} e^2_a \int d^2\bbox{k}_{T}\;
d_1\Bigl(x_A,(\bbox{k}_{T}+\case{1}{2}\bbox{q}_T)^2\Bigr)\;\bar{d}_2
\Bigl(x_B,(\bbox{k}_{T}-\case{1}{2}\bbox{q}_T)^2\Bigr) \\
&=&\case{1}{3}\sum_{a} e^2_a \int d^2\bbox{k}_{T}\;
d_1(x_A,\bbox{k}_{T}^2)\;\bar{d}_2 (x_B,\bbox{k}_{T}^2)+{\cal O}(Q_T^2)
\equiv I_0[d_1\bar{d}_2]+{\cal O}(Q_T^2),
\end{eqnarray}
assuming the distribution functions to be sufficiently well-behaved
to justify the Taylor expansion of the integrand.
For instance, for $V^{LL}_{2,2}$ of Eq.~(\ref{Dll}) one finds
in the limit $Q_T \rightarrow 0$ (omitting the flavor sum)
\begin{eqnarray}
&V^{LL}_{2,2}&\propto\int d^2\bbox{k}_{T}
\left(\case{1}{2} \bbox{k}_{T}^2 -\frac{(\bbox{q}_{T}\cdot\bbox{k}_{T})^2}
{Q_T^2}+\case{1}{2} Q_T^2\right)h^\perp_{1L}\Bigl(x_A,(\bbox{k}_{T}
+\case{1}{2}\bbox{q}_T)^2\Bigr)\,\bar{h}^\perp_{1L}
\Bigl(x_B,(\bbox{k}_{T}-\case{1}{2}\bbox{q}_T)^2\Bigr) \nonumber\\
&&=\frac{\bbox{q}_{T}^i\bbox{q}_{T}^j}{Q_T^2}\int d^2\bbox{k}_{T}
\left(\case{1}{2} \delta_{ij}\bbox{k}_{T}^2 -\bbox{k}_{Ti}\bbox{k}_{Tj}\right)
h^\perp_{1L}(x_A,\bbox{k}_{T}^2)\,\bar{h}^\perp_{1L}(x_B,\bbox{k}_{T}^2)\,
+{\cal O}(Q_T^2)={\cal O}(Q_T^2).
\label{tweedeorde}
\end{eqnarray}
So $V^{LL}_{2,2}$ has a kinematical zero of second order,
which is the natural behavior, because it multiplies a tensor that is
quadratic in $\hat{x}$ and $\hat{y}$. In the same fashion all structure
functions can be treated. Since this is a rather cumbersome
procedure, we will illustrate below the behavior by considering a Gaussian
$\bbox{k}_{T}^2$-dependence, from which the order of the kinematical zeros is
simply read off. We obtain only four structure functions {\em without\/}
a kinematical zero, in agreement with the results of RS,
\begin{mathletters}
\begin{eqnarray}
& & \left. W_T\right|_{Q_T=0} = I_0[f_1 \bar{f}_1],\\
& & \left. V^{LL}_T\right|_{Q_T=0} = -4I_0[g_{1L}\bar{g}_{1L}],\\
& & \left. V_{T}^{1\, TT}\right|_{Q_T=0} = I_0[\bbox{k}_{T}^2\frac{g_{1T}
\bar{g}_{1T}}{2M_A M_B}],\\
& & \left. \left(U^{TT}_{2,2} + \case{1}{2} U^{A\, TT}_{2,2}
+\case{1}{2} U^{B\, TT}_{2,2}\right)
\right|_{Q_T=0}= I_0[(h_{1T}+\frac{\bbox{k}_{T}^2}{2M_A^2}h_{1T}^\perp)
(\bar{h}_{1T}+\frac{\bbox{k}_{T}^2}{2M_B^2}\bar{h}_{1T}^\perp)].\label{argh}
\end{eqnarray}
\end{mathletters}
An easier way to obtain this result is to start from Eq.~(\ref{compact}),
and put $Q_T=0$ from there. One then never picks up the other structure
functions in the first place.
The cross sections at $Q_T=0$ can be obtained by insertion of these structure
functions into the explicit expressions given for the cross sections in
Eqs.~(\ref{unun})-(\ref{trtr}).

\subsection{Gaussian transverse momentum distributions}

It is instructive to consider a Gaussian $\bbox{k}_{T}$-dependence,
\begin{equation}
d (x,\bbox{k}_{T}^2) = d(x,0)\,\exp\left(-r^2\,\bbox{k}_{T}^2\right) ,
\end{equation}
where the transverse radius $r$ in principle depends on both the particular
distribution function, and on $x$. One can explicitly perform the convolution
integral~(\ref{Iconv}) (for simplicity we will omit the color
factor and flavor summation),
\begin{equation}
I[d_1\bar{d}_2]=\frac{\pi}{r_A^2+r_B^2}\exp\left( - Q_T^2\frac{r_A^2 r_B^2}
{r_A^2 + r_B^2} \right)\, d_1(x_A,0)\bar{d}_2(x_B,0),
\end{equation}
being regular in $Q_T = 0$. One can identify an `average' transverse
size $r$, given by $r^{-2}=r_A^{-2} + r_B^{-2}$.
We find for the structure functions,
\begin{mathletters}
\begin{eqnarray}
& &W_T = I[f_1 \bar{f}_1],\label{uu1}\\
& &V_T^{LL} = -4I[g_{1L}\bar{g}_{1L}],\label{Tll1}\\
& &V^{LL}_{2,2} =
\frac{8Q_T^2}{M_A M_B}\,\frac{r_A^2 r_B^2}{(r_A^2 + r_B^2)^2}\,
I[h_{1L}^\perp \bar{h}_{1L}^\perp],\label{Dll1}\\
& &V^{TL}_T = - \frac{Q_T}{M_A}\,\frac{r_B^2}{(r_A^2 + r_B^2)}\,
I[g_{1T}\bar{g}_{1L}],\\
& &V^{TL}_{2,2} = \frac{2Q_T^3}{M_A^2 M_B}\,
\frac{r_A^2 r_B^4}{(r_A^2 + r_B^2)^3}\,I[h_{1T}^\perp \bar{h}_{1L}^\perp]\\
& &U^{TL}_{2,2} = \frac{Q_T}{M_B}\,\frac{r_A^2}{(r_A^2 + r_B^2)}\,
I[h_{1T} \bar{h}_{1L}^\perp ]
+ \frac{Q_T}{2M_B}\,\frac{(r_A^2 - r_B^2)}{M_A^2 (r_A^2 + r_B^2)^2}\,
I[h_{1T}^\perp \bar{h}_{1L}^\perp],\\
& &V^{1\, TT}_T =
\frac{1}{2M_A M_B (r_A^2 + r_B^2)}\,I[ g_{1T}\bar{g}_{1T}],\\
& & V^{2\, TT}_T = -\frac{Q_T^2}{M_A M_B}\,\frac{r_A^2 r_B^2}{(r_A^2
+ r_B^2)^2}\,
I[g_{1T}\bar{g}_{1T}],\\
& &V^{2\, TT}_{2,2} = \frac{2Q_T^4}{M_A^2 M_B^2}\,
\frac{r_A^2 r_B^2}{(r_A^2 + r_B^2)^2}\,I[ h_{1T}^\perp
\bar{h}_{1T}^\perp ],\\
& & V^{1\, TT}_{2,2}+ U^{A\, TT}_{2,2}+U^{B\, TT}_{2,2}=
\frac{Q_T^2}{M_B^2}\,\frac{r_A^4}{(r_A^2 + r_B^2)^2}\,
I[h_{1T} \bar{h}_{1T}^\perp ]
+ \frac{Q_T^2}{M_A^2}\,\frac{r_B^4}{(r_A^2 + r_B^2)^2}\,
I[h_{1T}^\perp \bar{h}_{1T} ] \nonumber \\
& & \qquad \qquad
+\frac{Q_T^2}{2M_A^2 M_B^2}\,\frac{(r_A^4 - 4 r_A^2 r_B^2 + r_B^4)}
{(r_A^2 + r_B^2)^3}\,I[h_{1T}^\perp \bar{h}_{1T}^\perp],\\
& & U^{A\, TT}_{2,2}-U^{B\, TT}_{2,2}=
\frac{Q_T^2}{M_B^2}\,\frac{r_A^4}{(r_A^2 + r_B^2)^2}\,
I[h_{1T} \bar{h}_{1T}^\perp ]
- \frac{Q_T^2}{M_A^2}\,\frac{r_B^4}{(r_A^2 + r_B^2)^2}\,
I[h_{1T}^\perp \bar{h}_{1T} ] \nonumber \\
& & \qquad \qquad
+ \frac{Q_T^2}{2M_A^2 M_B^2}\,
\frac{(r_A^2 - r_B^2)} {(r_A^2 + r_B^2)^2}
\,I[h_{1T}^\perp \bar{h}_{1T}^\perp ],\\
& & U^{TT}_{2,2} + \case{1}{2} U^{A\, TT}_{2,2}
+\case{1}{2} U^{B\, TT}_{2,2}=
I[h_{1T} \bar{h}_{1T} ]
+ \frac{1}{2 M_A^2 (r_A^2 + r_B^2)}\,
\left( 1 + Q_T^2\,\frac{r_B^4} {r_A^2 + r_B^2}\right)
I[h_{1T}^\perp\bar{h}_{1T}] \nonumber \\
& &  \qquad \qquad
+ \frac{1}{2 M_B^2 (r_A^2 + r_B^2)}\,
\left( 1 + Q_T^2\,\frac{r_A^4} {r_A^2 + r_B^2}\right)
I[h_{1T}\bar{h}_{1T}^\perp ] \nonumber \\
& &  \qquad \qquad
+ \frac{1}{2M_A^2 M_B^2 (r_A^2 + r_B^2)^2}\,
\left( 1 + \frac{Q_T^2}{2}\,\frac{(r_A^2-r_B^2)^2} {r_A^2 + r_B^2}\right)
I[h_{1T}^\perp \bar{h}_{1T}^\perp ].
\end{eqnarray}
\end{mathletters}
These explicit results illustrate the kinematical zeros, and can be used
to obtain their order. For instance, Eq.~(\ref{Dll1}) is an illustration
of the result in Eq.~(\ref{tweedeorde}).
The expressions can also be used to illustrate the expected behavior of the
longitudinal-longitudinal asymmetry in Eq.~(\ref{llas}), which becomes
\begin{equation}
A_{\lambda_A\lambda_B}=\lambda_A\lambda_B\left[
\frac{I[g_{1L} \bar g_{1L}]}{I[f_1 \bar f_1 ]}
+\frac{\sin^2\theta\cos2\phi}{1+ \cos^2\theta}
\frac{Q_T^2}{M_A M_B}\,\frac{r_A^2 r_B^2}{(r_A^2 + r_B^2)^2}\,
\frac{I[h_{1L}^\perp \bar h_{1L}^\perp ]}{I[f_{1} \bar f_{1} ]}\right].
\end{equation}
The second (angle-dependent) term in the asymmetry starts off with
$Q_T^2$, and is proportional to the average transverse radius squared of
the functions $h_{1L}^\perp$ and $\bar h_{1L}^\perp$.
As a second example we note the structure function $V_T^{TL}$, being directly
proportional to the transverse radius squared of the helicity distribution
$\bar g_{1L}$ for hadron $B$.

Finally, we mention the CERN NA10 experiment~\cite{NA10}, which indicates
a possible $\sin^2 \theta\cos 2\phi$-asymmetry in {\em unpolarized\/} DY
scattering at measured $Q_T$, not suppressed by powers of $1/Q$.
Such an asymmetry, however, does not appear in the leading-order
result~(\ref{unun}).
It is, however, important to note that any kind of polarization in the beams
leads to such an asymmetry. Recently,
Brandenburg, Nachtmann, and Mirkes~\cite{bran93} have suggested that the
nontrivial QCD vacuum structure could be responsible for the asymmetry.
Another suggestion is that higher-twist effects are
responsible~\cite{bran94}.

\subsection{Conclusion}

In this paper we have investigated the leading-order polarized Drell-Yan
cross section at measured transverse momentum of the lepton pair
$Q_T\lesssim \Lambda$.
It is sensitive to the intrinsic transverse momentum of the quarks and
antiquarks in the colliding hadrons.
The results can be written in terms of six quark and six antiquark
distributions, depending on longitudinal lightcone momentum fraction
$x$, and transverse momentum $\bbox{k}_{T}^2$.
For each flavor they represent the lightcone momentum, helicity, and
transverse polarization distributions.
Measurements of these distributions require a study of the asymmetries
in doubly-polarized DY scattering, using longitudinally and transversely
polarized beams.

The quark distributions of Sec.~\ref{For} incorporate the intrinsic
$\bbox{k}_T$-dependence. Although this involves higher-twist operators,
the presence of a second scale $Q_T$ in the process enables one to find
observable effects that are not suppressed by the large scale $Q$.
Similarly, $\bbox{k}_T$-dependent correlation functions can be used to
analyze other deep inelastic processes, particularly semi-inclusive DIS.
They will then appear in convolutions with the quark fragmentation functions.
At this stage factorization has not been proven, so we do not know if the
distribution functions are universal.

\acknowledgements

We acknowledge numerous discussions with J. Levelt (Erlangen).
This work was supported by the Foundation for Fundamental Research on Matter
(FOM) and the National Organization for Scientific Research (NWO).

\appendix

\section{Perpendicular projection method}

Consider the following convolutions:
\begin{equation}
J[k_{1\perp}^{\mu_1}k_{2\perp}^{\mu_2}\ldots k_{n\perp}^{\mu_n}]=
\int d^2\bbox{k}_{aT}\,d^2\bbox{k}_{bT}\;\delta^2(\bbox{k}_{aT}+\bbox{k}_{bT}
-\bbox{q}_{T}) F(\bbox{k}_{aT}^2,\bbox{k}_{bT}^2)
k_{1\perp}^{\mu_1}k_{2\perp}^{\mu_2}\ldots k_{n\perp}^{\mu_n} .
\end{equation}
The perpendicular vectors $k_{i\perp}$ are taken from the set
$\{k_{a\perp},k_{b\perp}\}$.
In practise we only needed the cases where $n=0,1,2,3$, and $n=4$ with
two of the indices symmetrized.
We want to project these perpendicular tensors onto a basis of $XY$-tensors,
multiplied with scalar functions of $Q_T$.

Take the simplest non-trivial case of $n=1$,
\begin{equation}
J[k_{1\perp}^{\mu}]=\int d^2\bbox{k}_{aT}\,d^2\bbox{k}_{bT}\;\delta^2
(\bbox{k}_{aT}+\bbox{k}_{bT}-\bbox{q}_{T}) F(\bbox{k}_{aT}^2,\bbox{k}_{bT}^2)
k_{1\perp}^{\mu} .
\end{equation}
Since the only perpendicular direction available is
$q_\perp^\mu\approx X^\mu$,
this expression must be proportional to $\hat{x}^\mu$. The proportionality
constant is easily obtained by contracting with $\hat{x}_\mu$, and using
$\hat{x}^2=-1$.
We get (discarding corrections of ${\cal O}(1/Q^2)$ in this appendix)
\begin{equation}
J[k_{1\perp}^{\mu}]=-\hat{x}^\mu J[\hat{x}\cdot
k_{1\perp}]=\hat{x}^\mu J[\bbox{q}_{T}\cdot\bbox{k}_{1T}]/Q_T,
\end{equation}
making use of Eq.~(\ref{Pinprod}). We choose to write the integrand as a
function of $Q_T$, $\bbox{k}_{aT}^2$, and $\bbox{k}_{bT}^2$, by making use
of the relations
\begin{eqnarray}
\bbox{q}_{T}\cdot\bbox{k}_{aT}&=&\case{1}{2}(Q_T^2+\bbox{k}_{aT}^2
-\bbox{k}_{bT}^2),\\
\bbox{q}_{T}\cdot\bbox{k}_{bT}&=&\case{1}{2}(Q_T^2-\bbox{k}_{aT}^2
+\bbox{k}_{bT}^2),\\
\bbox{k}_{aT}\cdot\bbox{k}_{bT}&=&\case{1}{2}(Q_T^2-\bbox{k}_{aT}^2
-\bbox{k}_{bT}^2).
\end{eqnarray}
For the more intricate cases, $n>1$, the procedure is essentially the same.
Instead of $\hat{x}^\mu$, and $\hat{y}^\mu$, we will use the more convenient
building blocks $\hat{x}^\mu$, and $g_\perp^{\mu\nu}=-\hat{x}^\mu\hat{x}^\nu
-\hat{y}^\mu\hat{y}^\nu$. For $n=2$ we find
\begin{eqnarray}
J[k_{1\perp}^{\mu}k_{2\perp}^{\nu}]&=&
\left(\begin{array}{cc}\hat{x}^\mu\hat{x}^\nu,&g_\perp^{\mu\nu}\end{array}
\right)
\left(\begin{array}{cc} 2&1\\1&1\end{array}\right)
\left(\begin{array}{l} J[(\bbox{q}_{T}\cdot\bbox{k}_{1T})(\bbox{q}_{T}\cdot
\bbox{k}_{2T})]/Q_T^2\\
-J[\bbox{k}_{1T}\cdot\bbox{k}_{2T}]\end{array}\right)\\
&=&(2 \hat{x}^\mu\hat{x}^\nu+g_\perp^{\mu\nu})
J[(\bbox{q}_{T}\cdot\bbox{k}_{1T})(\bbox{q}_{T}\cdot\bbox{k}_{2T})]/Q_T^2-
(\hat{x}^\mu\hat{x}^\nu+g_\perp^{\mu\nu})J[\bbox{k}_{1T}\cdot\bbox{k}_{2T}],
\end{eqnarray}
where the matrix notation of the first line should be clear from the second.
In the same notation, the $n=3$-case reads
\begin{eqnarray}
\lefteqn{J[k_{1\perp}^{\mu}k_{2\perp}^{\nu}k_{3\perp}^\rho]=
\left(\begin{array}{cccc} \hat{x}^\mu\hat{x}^\nu\hat{x}^\rho, &
\hat{x}^\rho g_\perp^{\mu\nu}, & \hat{x}^\nu g_\perp^{\mu\rho} ,&
\hat{x}^\mu g_\perp^{\nu\rho} \end{array}\right)}\nonumber\\
&\qquad\qquad\qquad\qquad&\times\left(\begin{array}{cccc} 4&1&1&1\\1&1&0&0\\
1&0&1&0\\1&0&0&1\end{array}\right)
\left(\begin{array}{l} J[(\bbox{q}_{T}\cdot\bbox{k}_{1T})(\bbox{q}_{T}\cdot
\bbox{k}_{2T})(\bbox{q}_{T}\cdot\bbox{k}_{3T})]
/Q_T^3\\-J[(\bbox{q}_{T}\cdot\bbox{k}_{3T}) (\bbox{k}_{1T}\cdot\bbox{k}_{2T})]
/Q_T\\-J[(\bbox{q}_{T}\cdot\bbox{k}_{2T}) (\bbox{k}_{1T}\cdot\bbox{k}_{3T})]
/Q_T\\-J[(\bbox{q}_{T}\cdot\bbox{k}_{1T}) (\bbox{k}_{2T}\cdot\bbox{k}_{3T})]
/Q_T\end{array}\right).
\end{eqnarray}
Finally, for $n=4$ and two indices symmetrized, we find
\begin{eqnarray}
\lefteqn{J[k_{1\perp}^{\{\mu}k_{2\perp}^{\nu\}}k_{3\perp}^\rho
k_{4\perp}^\sigma]=
\left(\begin{array}{cccccc} \hat{x}^\mu\hat{x}^\nu\hat{x}^\rho\hat{x}^\sigma,
&\hat{x}^\mu\hat{x}^\nu g_\perp^{\rho\sigma},&\hat{x}^\rho\hat{x}^\sigma
g_\perp^{\mu\nu},&
\case{1}{2}\hat{x}^\rho\hat{x}^{\{\mu}g_\perp^{\nu\}\sigma},&
\case{1}{2}\hat{x}^\sigma\hat{x}^{\{\mu}g_\perp^{\nu\}\rho},&
g_\perp^{\mu\nu}g_\perp^{\rho\sigma}\end{array}\right)}\nonumber\\
&\qquad&\times\left(\begin{array}{cccccc} 8&2&2&2&2&1\\2&2&1&0&0&1\\
2&1&2&0&0&1\\2&0&0&2&0&0\\2&0&0&0&2&0\\1&1&1&0&0&1\end{array}\right)
\left(\begin{array}{l} 2J[(\bbox{q}_{T}\cdot\bbox{k}_{1T})(\bbox{q}_{T}\cdot
\bbox{k}_{2T})(\bbox{q}_{T}\cdot\bbox{k}_{3T})
(\bbox{q}_{T}\cdot\bbox{k}_{4T})]/Q_T^4 \\ -2J[(\bbox{q}_{T}
\cdot\bbox{k}_{1T})(\bbox{q}_{T}\cdot\bbox{k}_{2T})
(\bbox{k}_{3T}\cdot\bbox{k}_{4T})]/Q_T^2\\-2J[(\bbox{q}_{T}
\cdot\bbox{k}_{3T})(\bbox{q}_{T}\cdot\bbox{k}_{4T})
(\bbox{k}_{1T}\cdot\bbox{k}_{2T})]/Q_T^2\\-J[(\bbox{q}_{T}
\cdot\bbox{k}_{1T})(\bbox{q}_{T}\cdot\bbox{k}_{3T})
(\bbox{k}_{2T}\cdot\bbox{k}_{4T})+1\leftrightarrow 2]/Q_T^2\\
-J[(\bbox{q}_{T}\cdot\bbox{k}_{1T})(\bbox{q}_{T}\cdot\bbox{k}_{4T})
(\bbox{k}_{2T}\cdot\bbox{k}_{3T})+1\leftrightarrow 2]/Q_T^2\\
2J[(\bbox{k}_{1T}\cdot\bbox{k}_{2T})(\bbox{k}_{3T}\cdot\bbox{k}_{4T})]
\end{array}\right).
\end{eqnarray}
For the implementation of these rather lengthy formulas we used
FORM~\cite{verm91}.

\begin{figure}
\caption{\label{fig:handbag}Quark and antiquark handbag diagrams for
inclusive DIS.}
\end{figure}

\begin{figure}
\caption{\label{fig:DY}The quark and antiquark Born diagrams for the
Drell-Yan process.}
\end{figure}

\begin{figure}
\caption{\label{fig:blob}The blob representing the quark
correlation function.}
\end{figure}

\end{document}